\documentclass[preprint,3p,11pt, times, authoryear]{elsarticle}
 
\usepackage{booktabs}
\usepackage{float}

\usepackage{amsmath,amssymb,amsfonts,mathrsfs,hyperref,microtype, amsthm}

\usepackage[T1]{fontenc}

\usepackage{lmodern} 

\theoremstyle{plain}      

\newtheorem{lemma}{Lemma}[section]

\theoremstyle{definition} 

\newtheorem{remark}{Remark}[section]
\newtheorem*{exmp*}{Example}
\newcommand{\BS}{\boldsymbol}

\newcommand{\Rmnum}[1]{\expandafter\@slowromancap\romannumeral #1@}
\newcommand{\trp}{{\sf T}}
 \hypersetup{hidelinks}

\journal{Journal of }

\begin{document}
\begin{frontmatter}

\author[1]{Osama Idais\corref{cor1}}
\ead{osamaidais@gmail.com}

\author[2]{Eric Nyarko}
\ead{ericnyarko@ug.edu.gh}

\author[3]{Ola Alkhuffash}
\ead{ola.khuffash@yahoo.com}

\cortext[cor1]{Corresponding author}

\address[1]{Independent Researcher, Mannheim, Germany}
\address[2]{Department of Statistics and Actuarial Science, School of Physical and Mathematical Sciences, University of Ghana, Box LG 115, Legon, Accra, Ghana}
\address[3]{Department of Biostatistics, Faculty of Medicine, Lokman Hekim University, Ankara, Turkey}

\title{Regression Modeling for the Block–Basu Bivariate Exponential Distribution: Marginal and Conditional Means}

\begin{abstract}
The bivariate absolutely continuous exponential distribution introduced by \citet{block1974continuous} provides a model for dependent lifetimes in reliability and survival analysis. Its marginal means depend jointly and nonlinearly on the rate parameters, which complicates regression modeling and the interpretation of covariate effects on these means. In this paper, we develop two regression approaches based on a region-wise multivariate exponential-family representation. In the full-distribution approach, link functions relate the cumulant derivatives in a fixed reference region to linear predictors. These derivatives do not directly equal the marginal means, which are instead probability-weighted combinations of the regional derivatives. This relationship allows covariate effects on the marginal means to be examined. Likelihood derivatives are obtained through the chain rule, and the regression coefficients are estimated by maximum likelihood. The rate parameter estimates are then recovered through the inverse link functions and an explicit inverse mapping, with the admissible solution selected to ensure positive rates. As an alternative, we propose regression models based on the normalized conditional distribution within each region. In this setting, the cumulant derivatives coincide with the conditional means, allowing covariates to be linked directly to these means. Both approaches are evaluated through a simulation study.
\end{abstract}


\end{frontmatter}

\section{Introduction} \label{sec-1}

The theory of generalized linear models (GLMs) provides a standard framework for relating a response variable to explanatory variables by combining the three main components: the random component, the linear predictor and the link function (see \citet{nelder1972generalized, wedderburn1974quasi, mccullagh1983quasi, mccullagh1989generalized}).  The practical importance of GLMs lies in providing a unified methodology for responses following one-parameter exponential-family distributions, including for example the normal, binomial, Poisson, and gamma distributions. Therefore, applications of GLMs have arisen essentially in the univariate setting in many areas, such as insurance, healthcare and economics.

Although a general multivariate exponential-family representation exists (see \citet{FahrmeirTutz1994}), parametric multivariate distributions have different functional forms and dependence structures, which generally require distribution-specific formulations. In this situation, unlike the univariate GLM case, developing a unified modelling framework for multivariate responses remains challenging. Therefore, several extensions of the GLM have been proposed to accommodate such data structures. The approach of generalized estimating equations in its two versions; the first and second-order (GEE1 and GEE2) was developed to handle correlated binary data, where the correlations and dispersion allowed to vary with explanatory variables (see  \citet{liang1986longitudinal, smyth1989generalized, liang1992multivariate}). Later,  \citet{yee1996vector} developed vector generalized additive models whereas \citet{iwasaki2005bivariate} developed a GLM for a bivariate mixture distribution under the condition that bivariate responses are constantly correlated and are coming from a natural exponential family. This work was followed by further development  by \citet{neto2009bivariate} in the context of interval-valued data sets.   More recently, \citet{bonat2016multivariate} developed multivariate covariance GLMs that combine response-specific mean models with explicit covariance modelling. It provided a general framework for non-normal multivariate data analysis defined in terms of a covariance link function combined with a matrix linear predictor.  However, the use of GLMs continues to receive considerable attention in recent statistical research (see \citet{yousif2023comparative, lai2023extension, tsang2023post, correia2026verifying, knight2026fast, theus2026generalized, mathur2026parsimonious, javed2026logistic}). 

In this paper, we develop a GLM-based regression framework for bivariate observations following the three-parameter Block–Basu bivariate absolutely continuous exponential (BBBE) distribution derived by \citet{block1974continuous}. This is one of the most widely used lifetime distributions, obtained by conditioning the bivariate Marshall--Olkin model on unequal component lifetimes (see \citet{marshall1967multivariate}), thereby removing the singular component associated with simultaneous failures.  These distributions receive considerable attention in the competing risks analysis in reliability and survival analysis (see \citet{shen2018dependent, ragab2019prognostics, samanta2021bayesian, samanta2024competing, zhang2026inference, kumar2025robust}). 

Earlier work has incorporated covariates into Bayesian analyses of the BBBE distribution. \citet{wang2003bayesian} considered regression in a competing-risks setting, while \citet{dos2011bayesian} studied censored bivariate lifetimes in the presence of covariates. In these formulations, covariates enter through the rate parameters or functions of them. When the scientific objective is to examine covariate effects on expected lifetimes, a simple regression specification for the rate parameters does not generally provide an equally simple interpretation on the mean scale. This issue is particularly relevant for the BBBE distribution because its marginal means depend jointly and nonlinearly on the rate parameters. In the present paper, we therefore develop a formulation that makes the relationships between the linear predictors, the regional cumulant derivatives, the rate parameters, and the response means transparent. Within each of the regions $\{Y_1<Y_2\}$ and $\{Y_1>Y_2\}$, the cumulant derivatives provide the basis for this construction because they appear in both the likelihood score and the moment identities. In the full-distribution approach, link functions are specified for these derivatives in a fixed reference region, while both regions contribute to the likelihood. An explicit inverse mapping connects the fitted regression quantities to admissible rate parameters, and the marginal means are obtained from their probability-weighted regional representation. This construction supports likelihood inference and the examination of covariate effects on the marginal means within a single formulation. Also, we consider the normalized conditional distributions within each region. It turns out that their cumulant derivatives give explicitly conditional means so that mean-based link functions support a direct interpretation on the conditional-mean scale. The two approaches thus address complementary objectives including regression based on the full joint distribution, with marginal means obtained from the regional components, and direct regression on the conditional means.

The remainder of the paper is organized as follows. Section~\ref{sec-2} introduces the BBBE distribution and its main properties. Section~\ref{sec-3} develops its region-wise exponential-family representation and derives the corresponding moment identities. Section~\ref{sec-4} formulates the regression approach for the full joint distribution, while Section~\ref{sec-5} develops regression models based on the normalized conditional distributions. Section~\ref{sec-6} evaluates both approaches through a simulation study. Section~\ref{sec-8} concludes the paper with a brief discussion.

\section{The Block--Basu bivariate exponential distribution}\label{sec-2}

\cite{block1974continuous} proposed a flexible bivariate exponential model with three positive parameters
$\lambda_1>0, \lambda_2>0, \lambda_3>0$ with $\lambda:=\lambda_1+\lambda_2+\lambda_3$.  We write $\BS{\lambda}=(\lambda_1, \lambda_2, \lambda_3)^\trp$ and $(Y_1,Y_2)^\trp \sim \mathrm{BBBE}(\BS{\lambda})$.
The model is absolutely continuous on $(0,\infty)^2$ and is naturally expressed in two pieces on the triangular regions
\begin{equation}
R_1:=\{(y_1,y_2)^\trp: 0<y_1<y_2\},\qquad 
R_0:=\{(y_1,y_2)^\trp: 0<y_2<y_1\}. \label{regions}
\end{equation}
The BBBE density function with the parameters $\lambda_1,\lambda_2,\lambda_3 > 0$ is given by
\begin{equation}
P(y_1,y_2;\BS{\lambda}) =
\begin{cases}
\frac{\lambda_1 (\lambda_2 + \lambda_3)\lambda}{\lambda_1 + \lambda_2} \exp[-\lambda_1 y_1 - (\lambda_2 + \lambda_3)y_2], & (y_1,y_2)^\trp\in R_1, \\
\frac{\lambda_2 (\lambda_1 + \lambda_3)\lambda}{\lambda_1 + \lambda_2} \exp[-\lambda_2 y_2 - (\lambda_1 + \lambda_3)y_1], & (y_1,y_2)^\trp\in R_0 \label{eq:BBdist}
\end{cases}
\end{equation}
Since the distribution is continuous, $P(Y_1=Y_2)=0$.  Throughout we denote the regional-specific exponential forms by
\begin{align}
P_1(y_1,y_2;\BS{\lambda}) &=\frac{\lambda_1 (\lambda_2 + \lambda_3)\lambda}{\lambda_1 + \lambda_2} \exp[-\lambda_1 y_1 - (\lambda_2 + \lambda_3)y_2] \label{eq:BBdist1}\\
P_0(y_1,y_2;\BS{\lambda}) &=\frac{\lambda_2 (\lambda_1 + \lambda_3)\lambda}{\lambda_1 + \lambda_2} \exp[-\lambda_2 y_2 - (\lambda_1 + \lambda_3)y_1]\label{eq:BBdist0}
\end{align}

The first- and second-order moments are given in closed form. The marginal means of $Y_1$ and $Y_2$ receptively are
\begin{align}
\mathbb{E}(Y_1)
&=
\frac{1}{\lambda_1+\lambda_3}
+
\frac{\lambda_2\lambda_3}{\lambda(\lambda_1+\lambda_2)(\lambda_1+\lambda_3)},\label{eq:EY1_lam}\\
\mathbb{E}(Y_2)
&=
\frac{1}{\lambda_2+\lambda_3}
+
\frac{\lambda_1\lambda_3}{\lambda(\lambda_1+\lambda_2)(\lambda_2+\lambda_3)}.\label{eq:EY2_lam}
\end{align}
The marginal variances are
\begin{align}
\operatorname{Var}(Y_1)
&=
\frac{1}{(\lambda_1+\lambda_3)^2}
+
\frac{\lambda_3\lambda_2\big(2\lambda_1\lambda+\lambda_3\lambda_2\big)}
{\lambda^2(\lambda_1+\lambda_2)^2(\lambda_1+\lambda_3)^2},\label{eq:VarY1_lam}\\
\operatorname{Var}(Y_2)
&=
\frac{1}{(\lambda_2+\lambda_3)^2}
+
\frac{\lambda_3\lambda_1\big(2\lambda_2\lambda+\lambda_3\lambda_1\big)}
{\lambda^2(\lambda_1+\lambda_2)^2(\lambda_2+\lambda_3)^2}.\label{eq:VarY2_lam}
\end{align}
The covariance is
\begin{equation}
\operatorname{Cov}(Y_1,Y_2)
=
\frac{(\lambda_1^2+\lambda_2^2)\lambda_3\lambda+\lambda_1\lambda_2\lambda_3^2}
{\lambda^2(\lambda_1+\lambda_2)^2(\lambda_1+\lambda_3)(\lambda_2+\lambda_3)},\label{eq:Cov12_lam}
\end{equation}
and the correlation $\rho$ is the usual normalization:
\begin{equation}
\rho:=\operatorname{Corr}(Y_1,Y_2)
=
\frac{\operatorname{Cov}(Y_1,Y_2)}
{\sqrt{\operatorname{Var}(Y_1)\operatorname{Var}(Y_2)}}.\label{eq:Corr12_lam}
\end{equation}
The parameters $\lambda_1$ and $\lambda_2$ are rate parameters associated with the two responses $Y_1$ and $Y_2$, while $\lambda_3$ governs the deviation from independence. All three parameters contribute to the marginal distributions and the dependence structure. At the boundary of $\lambda_3$, i.e, $\lambda_3=0$, the joint density factorizes into two univariate exponential densities with rates $\lambda_1$ and $\lambda_2$ and hence, $Y_1$ and $Y_2$ become independent, and both the covariance  and the correlation $\rho$ vanish.

\subsection{The conditional joint distributions}
Let $(Y_1,Y_2)^\trp \sim \mathrm{BBBE}(\BS{\lambda})$ from \eqref{eq:BBdist}. Since its joint density has different forms over the two ordering regions $R_1$ and $R_0$, it is useful to derive the corresponding conditional joint distributions. These distributions provide the basis for the region-wise exponential-family representation developed later in this work.

Define the indicator function $\delta : (0,\infty)^2 \rightarrow \{0,1\}$ where
\begin{equation}
\delta(y_1,y_2) =
\begin{cases}
1, & (y_1,y_2)^\trp\in R_1, \\
0 & (y_1,y_2)^\trp\in R_0. \label{eq:delta}
\end{cases}
\end{equation}

 \begin{lemma}\label{lema_pi}
Consider  the indicator  $\delta$ given by (\ref{eq:delta}).   Integrating  the BBBE piecewise density (\ref{eq:BBdist}) over the regions $R_1$ and  $R_0$ defined in (\ref{regions}) yields respectively the region probabilities 
\begin{align} 
\pi_{1}(\boldsymbol{\lambda})& =P_{\BS{\lambda}}(\delta=1)=P_{\BS{\lambda}}(Y_2>Y_1)=\frac{\lambda_1}{\lambda_1+\lambda_2},\label{eq:pi1}\\
\pi_{0}(\boldsymbol{\lambda})& =P_{\BS{\lambda}}(\delta=0)=P_{\BS{\lambda}}(Y_1>Y_2)=\frac{\lambda_2}{\lambda_1+\lambda_2}.  \label{eq:pi2}
\end{align}
Here, $\pi_{1}(\boldsymbol{\lambda})+\pi_{0}(\boldsymbol{\lambda})=1$. The expectations of $\delta$ and $1-\delta$ are given by
\begin{align} 
\mathbb{E}(\delta)=\pi_{1}(\boldsymbol{\lambda}), \qquad \mathbb{E}(1-\delta)=\pi_{0}(\boldsymbol{\lambda}). \label{eq:pi3}
\end{align}
\end{lemma} 
\begin{proof} See \ref{proof:lema_pi}
\end{proof}

\begin{lemma} \label{lem:cond-dens}
Denote $\lambda_{13}=\lambda_1+\lambda_3$, $\lambda_{23}=\lambda_2+\lambda_3$ and $\lambda=\lambda_1+\lambda_2+\lambda_3$.  The conditional joint densities of  $(Y_1,Y_2)^\trp$ on the regions $R_1$ and $R_0$ are given respectively by $P_{R_1}(y_1,y_2; \BS{\lambda})$ and $P_{R_0}(y_1,y_2; \BS{\lambda})$, where 
\begin{align}
P_{R_1}(y_1,y_2; \BS{\lambda})=P(y_1,y_2\mid Y_2>Y_1)
&=\lambda_{23}\lambda\,
\exp\!\big[-\lambda_1 y_1-\lambda_{23}y_2\big],\,\,  (y_1,y_2)^\trp\in R_1,\label{eq:cond_y2gt}\\
P_{R_0}(y_1,y_2; \BS{\lambda})=P(y_1,y_2\mid Y_1>Y_2)
&=\lambda_{13}\lambda\,
\exp\!\big[-\lambda_2 y_2-\lambda_{13}y_1\big]
,\,\,  (y_1,y_2)^\trp\in R_0.\label{eq:cond_y1gt}
\end{align}
\end{lemma}
\begin{proof}
Given the BBBE density (\ref{eq:BBdist}), by definition the conditional joint densities on each region, 
\[
P(y_1,y_2\mid Y_2>Y_1)=\frac{P(y_1,y_2; \BS{\lambda})}{P(Y_2>Y_1)},\qquad (y_1,y_2)^\trp\in R_1,
\]
and
\[
P(y_1,y_2\mid Y_1>Y_2)=\frac{P(y_1,y_2; \BS{\lambda})}{P(Y_1>Y_2)},\qquad  (y_1,y_2)^\trp\in R_0.
\]
Substituting \eqref{eq:BBdist}, \eqref{eq:pi1} and \eqref{eq:pi2}  and the above region probabilities gives \eqref{eq:cond_y2gt} and \eqref{eq:cond_y1gt},
 \end{proof}

\begin{remark}
In the conditional joint distributions (\ref{eq:cond_y2gt}) and (\ref{eq:cond_y1gt}), the parameter $\lambda_3$ is no longer characterizing independence. Although setting $\lambda_3=0$ yields independent exponential random variables in the unconditional BBBE distribution (\ref{eq:BBdist}), conditional independence does not hold within either region $\{Y_1<Y_2\}$ or $\{Y_2<Y_1\}$.  This is because the regional constraint itself induces dependence between $Y_1$ and $Y_2$, as the value of one variable restricts the possible values of the other. Therefore, $\lambda_3=0$ does not imply independence for the region-wise conditional joint distributions.
\end{remark}

\begin{lemma}\label{cond-moments}
Consider the conditional joint densities \eqref{eq:cond_y2gt} and \eqref{eq:cond_y1gt}. The conditional means, variances, and covariances of $Y_1$ and $Y_2$ are given by
\begin{align}
\mathbb E_{R_1}(Y_1)&=\dfrac{1}{\lambda},&{} \mathbb  E_{R_1}(Y_2)&=\dfrac{1}{\lambda}+\dfrac{1}{\lambda_{23}}\\
\mathrm{Var}_{R_1}(Y_1)&=\frac{1}{\lambda^2},&{} \mathrm{Var}_{R_1}(Y_2)&=\frac{1}{\lambda^2}+\frac{1}{\lambda_{23}^2}\\
\mathbb E_{R_0}(Y_1)&=\dfrac{1}{\lambda}+\dfrac{1}{\lambda_{13}}, &{}\mathbb E_{R_0}(Y_2)&=\dfrac{1}{\lambda} \\ 
\mathrm{Var}_{R_0}(Y_1)&=\frac{1}{\lambda^2}+\frac{1}{\lambda_{13}^2},&{}  \mathrm{Var}_{R_0}(Y_2)&=\frac{1}{\lambda^2}  \\
\mathrm{Cov}_{R_1}(Y_1,Y_2)&=\mathrm{Cov}_{R_0}(Y_1,Y_2)=\frac{1}{\lambda^2}.&&{}&&
\end{align}
\end{lemma}
\begin{proof} See \ref{proof:cond-moments}
\end{proof}

\begin{lemma}\label{cond-tot-moments}
Given $\pi_{1}(\boldsymbol{\lambda})$ and $\pi_{0}(\boldsymbol{\lambda})$ from Lemma \ref{lema_pi}. The total expectation $\mathbb{E}(Y_k)$, $k=1,2$, from  \eqref{eq:EY1_lam} and \eqref{eq:EY2_lam}, can be obtained as
\begin{align}
\mathbb{E}(Y_k)
&=\pi_{1}(\boldsymbol{\lambda})\,\mathbb{E}_{R_1}(Y_k)+\pi_{0}(\boldsymbol{\lambda})\,\mathbb{E}_{R_0}(Y_k)\,, k=1,2.
\end{align}
\end{lemma}

\begin{proof}
For $k=1,2$,
\begin{align*}
 \pi_{1}(\boldsymbol{\lambda})\,\mathbb{E}_{R_1}(Y_k)
&= \pi_{1}(\boldsymbol{\lambda})\int_0^\infty\int_{y_1}^\infty y_k\,P_{R_1}(y_1,y_2; \BS{\lambda}) dy_2\,dy_1\\
&=\pi_{1}(\boldsymbol{\lambda})\int_0^\infty\int_{y_1}^\infty y_k\,\frac{P(y_1,y_2; \BS{\lambda})}{P(Y_2>Y_1)}dy_2\,dy_1=\int_0^\infty\int_{y_1}^\infty y_k\,P(y_1,y_2; \BS{\lambda})dy_2\,dy_1
\end{align*}
Similarly, $\pi_{0}(\boldsymbol{\lambda})\,\mathbb{E}_{R_0}(Y_k)=\int_0^\infty\int_{y_2}^\infty y_k\,P(y_1,y_2; \BS{\lambda})dy_1\,dy_2,\,k=1,2$.
\end{proof}

\begin{remark}\label{rem: partial expectation}
In Lemma \ref{cond-tot-moments}, the quantity $\pi_{\delta}(\boldsymbol{\lambda})\,\mathbb{E}_{R_\delta}(Y_k)$, for any $k=1,2$, and $\delta=0,1$ represents the partial expectation or an unnormalized truncated mean over $R_\delta,\, \delta= 0,1$, because it presents the contribution of the corresponding region to the marginal expectation $\mathbb{E}(Y_k)$ on the whole region $R_1 \cup R_0$. The total expectation 
$\mathbb{E}(Y_k)$ is simply the sum of the partial expectations of non-overlapping subregions $R_1$ and $R_0$.
\end{remark}

\section{Exponential-family-like representation and properties} \label{sec-3}
In literature, a general form of Multivariate Exponential family is present, see \citet{mccullagh1989generalized,FahrmeirTutz1994}, in the form
 
\begin{equation}
P(\boldsymbol{y};\boldsymbol{\theta},\phi)
=
\exp\!\left\{
\Big(\boldsymbol{y}^\trp\boldsymbol{\theta}-b(\boldsymbol{\theta})\Big)/\phi+c(\boldsymbol{y},\phi)
\right\} \label{MVEF}
\end{equation}
To rewrite the BBBE density \eqref{eq:BBdist} in an exponential-family-like form (\ref{MVEF}), it is convenient to introduce the canonical parameters
\[
\theta_1:=-\lambda_1,\qquad 
\theta_2:=-\lambda_2,\qquad 
\theta_3:=-\lambda_3,
\qquad 
\theta:=\theta_1+\theta_2+\theta_3,
\]
so that \(\theta_1<0\), \(\theta_2<0\), \(\theta_3<0\), and \(\theta<0\). Also, let $\boldsymbol{\theta}=(\theta_1,\theta_2,\theta_3)^\trp$. Define the deterministic quantity $Y_3:=\delta Y_2+(1-\delta)Y_1=\max(Y_1,Y_2)$. This does not add a third response variable, that is  $Y_3=Y_1$ for $\delta=0$ and $Y_3=Y_2$ for $\delta=1$, where $\delta$ defined in (\ref{eq:delta}).  Define the statistic vector $\mathbf{T}(Y_1,Y_2):=(Y_1,Y_2,Y_3)^\trp,\,\,(Y_1,Y_2)^\trp\in(0,\infty)^2$.  Then the exponent in the BBBE density can be written in a single linear expression in three parameters, whereas the associate parameter $\lambda_3$ behaves as the rate parameter of $Y_3$.  Therefore, on the region determined by \(\delta\), the BBBE density admits the following exponential-family-like form
\begin{align}
P_{\delta}(\BS{y};\boldsymbol{\theta})
&=
\exp\!\left\{
\boldsymbol{\theta}^\trp \mathbf{T}(y_1,y_2)-b_\delta(\boldsymbol{\theta})
\right\}, \label{exp-Biv}\\
&=\exp\!\left\{\theta_1 y_1+\theta_2 y_2+\theta_3 y_3-b_\delta(\boldsymbol{\theta})\right\},  \label{eq:bb-expfam-theta}
\end{align}
where $\BS{y}=(y_1,y_2)^\trp \in R_\delta$, $\delta \in \{1,0\}$. The region-based cumulant function $b_\delta(.)$ is given by
\[
b_\delta(\boldsymbol{\theta})
=
\delta\, b_1(\boldsymbol{\theta})+(1-\delta)\, b_0(\boldsymbol{\theta}),\,\,\delta\in \{1,0\}
\]
where 
\begin{align}
b_1(\boldsymbol{\theta})
&=
-\log(-\theta)-\log(-\theta_1)-\log\!\bigl(-(\theta_2+\theta_3)\bigr)
+\log\!\bigl(-(\theta_1+\theta_2)\bigr),
\label{eq:b1-theta}\\
b_0(\boldsymbol{\theta})
&=
-\log(-\theta)-\log(-\theta_2)-\log\!\bigl(-(\theta_1+\theta_3)\bigr)
+\log\!\bigl(-(\theta_1+\theta_2)\bigr).
\label{eq:b0-theta}
\end{align}

In the following, we state the essential regularity conditions that permit differentiation under the integral sign; see \cite{dobson2018introduction} and \cite{mccullagh1989generalized}. Under the exponential-family-like representation \eqref{exp-Biv}, we derive the moment identities using an argument analogous to that for a standard exponential family, while accounting for the piecewise structure of the BBBE density over the two regions $R_1$ and $R_0$. The region-wise representations \eqref{eq:BBdist1} and \eqref{eq:BBdist0} in canonical form satisfy
\begin{equation}
\begin{aligned}
1
&=
\iint_{R_1}
P_1(\BS{y};\boldsymbol{\theta})\,dy_1\,dy_2
+
\iint_{R_0}
P_0(\BS{y};\boldsymbol{\theta})\,dy_1\,dy_2 .
\end{aligned}
\label{eq:norm-split}
\end{equation}

Differentiating \eqref{eq:norm-split} with respect to $\theta_j$ and
interchanging differentiation and integration give
\begin{equation}
\iint_{R_1}
\frac{\partial P_1(\BS{y};\boldsymbol{\theta})}
     {\partial\theta_j}\,dy_1\,dy_2
+
\iint_{R_0}
\frac{\partial P_0(\BS{y};\boldsymbol{\theta})}
     {\partial\theta_j}\,dy_1\,dy_2
=
0,
\qquad j=1,2,3.
\label{eq:first-diff}
\end{equation}

Differentiating once more with respect to $\theta_k$ yields
\begin{equation}
\iint_{R_1}
\frac{\partial^2P_1(\BS{y};\boldsymbol{\theta})}
     {\partial\theta_j\partial\theta_k}\,dy_1\,dy_2
+
\iint_{R_0}
\frac{\partial^2P_0(\BS{y};\boldsymbol{\theta})}
     {\partial\theta_j\partial\theta_k}\,dy_1\,dy_2
=
0,
\qquad j,k=1,2,3.
\label{eq:second-diff}
\end{equation}

Equations \eqref{eq:first-diff} and \eqref{eq:second-diff} provide the
basis for deriving the total mean vector and covariance matrix of the BBBE distribution. The following lemma is immediate.

\begin{lemma}\label{marg-moments}
Let $\mathbf{T}(Y_1,Y_2)=(Y_1,Y_2,Y_3)^\trp$, $Y_3=\max(Y_1,Y_2)$. Let $\BS{\theta}=(\theta_1,\theta_2,\theta_3)^\trp$ and from Lemma \ref{lema_pi}, denote $\mathbb{E}(\delta)=\pi_{1}(\BS{\theta})=\theta_1/(\theta_1+\theta_2)$. 
The total mean vector of $\mathbf{T}(Y_1,Y_2)$ is
\begin{equation}
\mathbb{E}\!\left[\mathbf{T}(Y_1,Y_2)\right]
=
\mathbb{E}(\delta)\nabla b_1(\boldsymbol{\theta})
+
\mathbb{E}(1-\delta)\nabla b_0(\boldsymbol{\theta}).
\label{eq:mean-vector-final}
\end{equation}
Moreover, its covariance matrix is
\begin{align}
\operatorname{Var}\!\left\{\mathbf{T}(Y_1,Y_2)\right\}
={}&
\mathbb{E}(\delta)\nabla^2b_1(\boldsymbol{\theta})
+
\mathbb{E}(1-\delta)\nabla^2b_0(\boldsymbol{\theta})
\notag\\
&+
\mathbb{E}(\delta)\mathbb{E}(1-\delta)
\left\{
\nabla b_1(\boldsymbol{\theta})
-
\nabla b_0(\boldsymbol{\theta})
\right\}
\notag\\
&\hspace{3.5cm}\times
\left\{
\nabla b_1(\boldsymbol{\theta})
-
\nabla b_0(\boldsymbol{\theta})
\right\}^{\trp}
\notag\\
&+
\nabla\pi_1(\boldsymbol{\theta})
\left\{
\nabla b_1(\boldsymbol{\theta})
-
\nabla b_0(\boldsymbol{\theta})
\right\}^{\trp}
\notag\\
&+
\left\{
\nabla b_1(\boldsymbol{\theta})
-
\nabla b_0(\boldsymbol{\theta})
\right\}
\left\{\nabla\pi_1(\boldsymbol{\theta})\right\}^{\trp}.
\label{eq:cov-matrix-final}
\end{align}
\end{lemma}

\begin{proof} see \ref{proof:marg-moments}
\end{proof}
Thus, the total covariance is determined by the weighted
Hessian matrices, the difference between the two gradient vectors,
and the parameter dependence of the region probabilities.
The last two terms arise because the score and the ordering
indicator are generally correlated.

\section{A GLM formulation for the BBBE distribution} \label{sec-4}

Following the classical generalized linear model (GLM) framework of
\cite{nelder1972generalized}, we distinguish a random component, a
systematic component, and a link structure. The
random component belongs to an exponential family distribution,  the systematic
component represents the linear predictor, and estimation is
based on score equations and Fisher scoring. For the BBBE distribution, a similar construction can be
developed, with the essential difference that the density has two
region-specific representations determined by the ordering of the two
responses.
 
\subsection{Within-region GLM structure}

Let $(Y_{1i},Y_{2i})^\trp,\,i=1,\ldots,n$, be independent bivariate continuous observations from the BBBE
distribution. It follows that $Y_{3i}:=\max(Y_{1i},Y_{2i})$, $\delta_i:=I(Y_{1i}<Y_{2i})$, with the canonical parameters
$\theta_{1i}:=-\lambda_{1i}, \theta_{2i}:=-\lambda_{2i}, \theta_{3i}:=-\lambda_{3i}$, and $\boldsymbol{\theta}_i
= (\theta_{1i},\theta_{2i},\theta_{3i})^\trp, \theta_i=\theta_{1i}+\theta_{2i}+\theta_{3i}$. Since $\lambda_{ji}>0$, we have $\theta_{ji}<0$, $j=1,2,3$, and
$\theta_i<0$.  For $\mathbf{T}_i=(Y_{1i},Y_{2i},Y_{3i})^\trp$ and the observed region $\delta_i$, the corresponding exponential-family-like form is $P_{\delta_i}(\mathbf{T}_i;\boldsymbol{\theta}_i)=\exp
\left\{
\boldsymbol{\theta}_i^\trp\mathbf{T}_i
-
b_{\delta_i}(\boldsymbol{\theta}_i)
\right\}$,
where $
b_{\delta_i}(\boldsymbol{\theta}_i)
=
\delta_i b_1(\boldsymbol{\theta}_i)
+
(1-\delta_i)b_0(\boldsymbol{\theta}_i)$.  Denote $r=1$ when $\delta_i=1$ and otherwise $r=0$. For response component $j=1,2,3$ and region $r\in\{0,1\}$, define
the region-based cumulant  derivatives 
\begin{equation}
m_{jri}
=
\frac{\partial b_r(\boldsymbol{\theta}_i)}
     {\partial\theta_{ji}},\,\, \boldsymbol m_{ri}=(m_{1ri},m_{2ri},m_{3ri})^\trp=\nabla b_r(\boldsymbol\theta_i).
\label{eq:within_region_mean}
\end{equation}
Explicitly, the corresponding derivatives are
\begin{align}
m_{11i}=\frac{\partial b_1}{\partial\theta_{1i}}
&=
-\frac{1}{\theta_i}
-\frac{1}{\theta_{1i}}
+\frac{1}{\theta_{1i}+\theta_{2i}},
&
m_{10i}=\frac{\partial b_0}{\partial\theta_{1i}}
&=
-\frac{1}{\theta_i}
-\frac{1}{\theta_{1i}+\theta_{3i}}
+\frac{1}{\theta_{1i}+\theta_{2i}},
\label{eq:m1_regions}
\\
m_{21i}=\frac{\partial b_1}{\partial\theta_{2i}}
&=
-\frac{1}{\theta_i}
-\frac{1}{\theta_{2i}+\theta_{3i}}
+\frac{1}{\theta_{1i}+\theta_{2i}},
&
m_{20i}=\frac{\partial b_0}{\partial\theta_{2i}}
&=
-\frac{1}{\theta_i}
-\frac{1}{\theta_{2i}}
+\frac{1}{\theta_{1i}+\theta_{2i}},
\label{eq:m2_regions}
\\
m_{31i}=\frac{\partial b_1}{\partial\theta_{3i}}
&=
-\frac{1}{\theta_i}
-\frac{1}{\theta_{2i}+\theta_{3i}},
&
m_{30i}=\frac{\partial b_0}{\partial\theta_{3i}}
&=
-\frac{1}{\theta_i}
-\frac{1}{\theta_{1i}+\theta_{3i}}.
\label{eq:m3_regions}
\end{align}

\begin{remark}
The region-based cumulant $b_r$ contains the region
probability, therefore the quantities $m_{jr}$  are not the conditional means $\mathbb E_{R_r}(Y_j)$ of
Lemma~\ref{cond-moments}; for example $m_{11}=1/\lambda+1/\lambda_{1}-1/(\lambda_{1}+\lambda_{2})$,
whereas $\mathbb E_{R_1}(Y_1)=1/\lambda$.  This can be seen in the normalized conditional joint density whose cumulant function
$b_r+\log\pi_{r}$ and thus $\mathbb E_{R_{r}}(Y_{j})=m_{jr}+\partial \log \pi_{r}/\partial\theta_{j}$ with $\pi_{1}=\theta_{1}/(\theta_{1}+\theta_{2})$ and
$\pi_{0}=\theta_{2}/(\theta_{1}+\theta_{2})$. However, by Lemma~\ref{marg-moments}, $\boldsymbol\mu=\mathbb E(\mathbf T)=\pi_{1}\boldsymbol m_{1}+\pi_{0}\boldsymbol m_{0}$.
\end{remark}

In the classical GLM, the response mean $\mu_i=b'(\theta_i)$ connects to the natural parameter by the link function $g(\mu_i)=\eta_i$, see \citet{dobson2018introduction}. Here, the regional derivatives appear directly in the likelihood score. This motivates specifying link functions on these quantities $m_{jri}$, while obtaining the marginal means from the resulting full distribution as described in Lemma \ref{marg-moments}. We use region $r=1$ as the anchor region to specify the regression; choosing
$r=0$ gives an analogous construction. For $j=1,2,3$, let
\begin{equation}
\eta_{ji}
=
g_{j}(m_{j1i})
=
\mathbf f_{j}(\mathbf x_i)^\trp\boldsymbol\beta_{j},
\qquad
m_{j1i}=g_{j}^{-1}(\eta_{ji}),
\label{eq:regional_predictor}
\end{equation}
where $g_{j}$ is the link function, $\eta_{ji}$ is the linear predictor, $\mathbf f_{j}(\mathbf x_i)=\bigl(1,f_{j1}(\mathbf x_i),\ldots,f_{jk_{j}}(\mathbf x_i)\bigr)^\trp$,
and $\boldsymbol\beta_{j}=(\beta_{j0},\ldots,\beta_{jk_{j}})^\trp$.  The complete $p$-dimensional coefficient vector is
\begin{equation}
\boldsymbol\beta
=
\left(
\boldsymbol\beta_{1}^\trp,
\boldsymbol\beta_{2}^\trp,
\boldsymbol\beta_{3}^\trp
\right)^\trp,
\qquad
p=\sum_{j=1}^{3}(k_{j}+1).
\label{eq:complete_beta}
\end{equation}
In this formulation, we propose a general formulation of the regression models  to cover heterogeneity  between components.
Each component may have its own link function, regression functions, and regression parameters. This allows the covariates, functional forms, and dimensions of the regression predictors to vary across components. Models with shared links, common predictors, or equal regression parameters are included as restricted special cases. 
The linked quantities must give a unique admissible solution
of $\boldsymbol m_{1i}=\nabla b_1(\boldsymbol\theta_i)$. Define the inverse $H_r=(\nabla b_r)^{-1}$ for $r=1,0$.  We obtain $\boldsymbol\theta_i=H_1(\boldsymbol m_{1i})$, i.e., 
\begin{equation}
\boldsymbol\theta_i
=
H_1\left(
\begin{pmatrix}
g_{1}^{-1}(\eta_{1i})\\
g_{2}^{-1}(\eta_{2i})\\
g_{3}^{-1}(\eta_{3i})
\end{pmatrix}
\right).
\label{eq:theta_from_predictors}
\end{equation}
Thus, $\boldsymbol\theta_i$ is a function of
$\mathbf x_i$ and $\boldsymbol\beta$ and is not estimated
separately. The quantities for region $r=0$ are then
$\boldsymbol m_{0i}=\nabla b_0(\boldsymbol\theta_i)=\nabla b_0\{H_1(\boldsymbol m_{1i})\}$. Hence $H_1(\boldsymbol m_{1i})=H_0(\boldsymbol m_{0i})=\boldsymbol\theta_i$. 
No additional region-$0$ coefficient vectors are introduced.
Both regions are nevertheless used in the full likelihood.  From \eqref{eq:m1_regions}--\eqref{eq:m3_regions},
\begin{equation}
m_{21i}-m_{31i}=m_{10i}-m_{30i}=\frac1{\theta_{1i}+\theta_{2i}}<0 .
\label{eq:compat_identity}
\end{equation}

Write $A_{1i}=m_{21i}-m_{31i}$, $C_{1i}=m_{11i}-A_{1i}$, $D_{1i}=m_{31i}$ and $q_i=1/\theta_i$.
By \eqref{eq:m1_regions}--\eqref{eq:compat_identity}, $A_{1i}=1/(\theta_{1i}+\theta_{2i})$,
$C_{1i}=-q_i-1/\theta_{1i}$, $D_{1i}=-q_i-1/(\theta_{2i}+\theta_{3i})$, so
$\theta_{1i}=-1/(C_{1i}+q_i)$ and $\theta_{2i}+\theta_{3i}=-1/(D_{1i}+q_i)$. The identity
$\theta_i=\theta_{1i}+(\theta_{2i}+\theta_{3i})$ reads
$1/q_i=-1/(C_{1i}+q_i)-1/(D_{1i}+q_i)$, i.e.
\begin{equation}
3q_i^2+2(C_{1i}+D_{1i})q_i+C_{1i}D_{1i}=0,\qquad
q_i=\frac{-(C_{1i}+D_{1i})\pm\sqrt{C_{1i}^2-C_{1i}D_{1i}+D_{1i}^2}}{3}.
\label{eq:q_region1}
\end{equation}
The root is chosen with $q_i<0$, $C_{1i}+q_i\neq0$, $D_{1i}+q_i\neq0$ and
$\boldsymbol\theta_i$ in the admissible space ($\theta_{ji}<0$, $j=1,2,3$). Then
\begin{equation}
\theta_{1i}=-\frac1{C_{1i}+q_i},\qquad
\theta_{2i}=\frac1{A_{1i}}-\theta_{1i},\qquad
\theta_{3i}=\frac1{q_i}-\frac1{A_{1i}} .
\end{equation}
For the region $r=0$, put $A_{0i}=m_{10i}-m_{30i}=1/(\theta_{1i}+\theta_{2i})$,
$C_{0i}=m_{20i}-A_{0i}$, $D_{0i}=m_{30i}$; the same quadratic holds with $C_{0i},D_{0i}$, and
$\theta_{2i}=-1/(C_{0i}+q_i)$, $\theta_{1i}=1/A_{0i}-\theta_{2i}$, $\theta_{3i}=1/q_i-1/A_{0i}$.

The marginal means are then induced following Lemma~\ref{marg-moments}, with $\boldsymbol\theta_i=\boldsymbol\theta_i(\boldsymbol\beta)$ from
\eqref{eq:theta_from_predictors},
\begin{equation}
\boldsymbol\mu_i=\mathbb E(\mathbf T_i)=\pi_{1i}\,\boldsymbol m_{1i}+\pi_{0i}\,\nabla b_0\{H_1(\boldsymbol m_{1i})\},
\qquad \pi_{1i}=\frac{\theta_{1i}}{\theta_{1i}+\theta_{2i}},\quad \pi_{0i}=1-\pi_{1i}.
\label{eq:marginal_mean_vector}
\end{equation}
Thus $\mu_{ji}(\boldsymbol\beta)=\pi_{1i}(\boldsymbol\beta)\,g_{j}^{-1}\{\mathbf f_{j}(\mathbf x_i)^\trp\boldsymbol\beta_{j}\}
+\pi_{0i}(\boldsymbol\beta)\,\partial b_0(\BS{\theta(\BS{\beta})})/\partial\theta_{ji}$ evaluated at $\boldsymbol\theta_i=H_1(\boldsymbol m_{1i})$;
the weights depend on $\boldsymbol\beta$ through $\boldsymbol\theta_i$ only.

\begin{remark}
For the log link, $m_{j1i}=e^{\eta_{ji}}$, whereas the reciprocal link gives $m_{j1i}=1/\eta_{ji}$, $\eta_{ji}\ne0$. The log link requires all three linked quantities to be positive. This restricts the BBBE parameter space because
$m_{21i}$ need not be positive. Neither inverse link alone
guarantees an admissible $\boldsymbol\theta_i$; the restrictions must also hold. 
The canonical link is not componentwise in $m_{jri}$. It sets
$\theta_{ji}=\mathbf f_j(\mathbf x_i)^\trp\boldsymbol\beta_j$, $j=1,2,3$, with
$\theta_{ji}<0$ over the design space. The inverse mapping $H_r$  is not needed. 
\end{remark}

The approach links the covariates to regional cumulant derivatives $m_{j1}(x)$ and not  directly to the marginal means. Therefore, a single regression coefficient does not usually give the effect of the covariate on the marginal mean. This effect must be calculated from the fitted marginal means $\mu_j(x)$. For a binary
covariate, we can use the difference $\mu_j(1)-\mu_j(0)$ or the ratio $\mu_j(1)/\mu_j(0)$. For a continuous covariate, we can use the
derivative $\partial\mu_j(x)/\partial x$, or we can compare the fitted means at two chosen values of the covariate, while the other covariates
are kept fixed. Standard errors and confidence intervals for these quantities are obtained by the delta method, using the inverse of the
Fisher information $\mathcal{I}(\beta)$.

\subsection{Scores and Fisher information}

The log-likelihood contribution of observation $i$ is
\begin{equation}
\ell_i(\boldsymbol\theta_i(\BS{\beta});\delta_i)=\sum_{j=1}^3\theta_{ji}(\BS{\beta})y_{ji}-b_{\delta_i}(\boldsymbol\theta_i(\BS{\beta})),
\label{eq:bb_loglik_i}
\end{equation}
and $\ell(\boldsymbol\beta)=\sum_{i=1}^n\ell_i\{\boldsymbol\theta_i(\boldsymbol\beta);\delta_i\}$. 
Since $m_{j\delta_ii}=\partial b_{\delta_i}/\partial\theta_{ji}$,
\begin{equation}
\frac{\partial\ell_i}{\partial\theta_{ji}}=y_{ji}-m_{j\delta_ii},\qquad j=1,2,3.
\label{eq:dl_dtheta}
\end{equation}
Each $\theta_{ji}$ depends on all three  $m_{\ell1i}, \ell=1,2,3$ through $H_1$, while $\eta_{\ell i}$ depends
only on $\boldsymbol\beta_{\ell}$. Hence, for $k,j=1,2,3$,
\begin{equation}
\frac{\partial\theta_{ki}}{\partial\boldsymbol\beta_{j}}
=\sum_{\ell=1}^3\frac{\partial\theta_{ki}}{\partial m_{\ell1i}}
\frac{\partial m_{\ell1i}}{\partial\eta_{\ell i}}
\frac{\partial\eta_{\ell i}}{\partial\boldsymbol\beta_{j}}
=\frac{\partial\theta_{ki}}{\partial m_{j1i}}\;
\frac1{g_{j}'(m_{j1i})}\;\mathbf f_{j}(\mathbf x_i).
\label{eq:score_chain}
\end{equation}
Here, only the term $\ell=j$ survives because $\eta_{\ell i}$ depends on $\BS{\beta}_{\ell }$. The factor $\partial\theta_{ki}/\partial m_{j1i}$ is
\emph{not} the reciprocal of $\partial m_{j1i}/\partial\theta_{ki}$ for a single pair.
Since $\boldsymbol m_{1i}=\nabla b_1(\boldsymbol\theta_i)$, we have
$\partial\boldsymbol m_{1i}/\partial\boldsymbol\theta_i^\trp=\nabla^2b_1(\boldsymbol\theta_i)$, and 
\begin{equation}
\partial\boldsymbol\theta_i/\partial\boldsymbol m_{1i}^\trp=
\Bigl(\frac{\partial\theta_{ki}}{\partial m_{\ell1i}}\Bigr)_{k,\ell=1}^3=\{\nabla^2b_1(\boldsymbol\theta_i)\}^{-1}.
\label{eq:dtheta_dm}
\end{equation}
Denote by $[\nabla^2b_1(\boldsymbol\theta_i)]^{-1}_{kj}$ the
$(k,j)$ element of this inverse.
Combining \eqref{eq:dl_dtheta}--\eqref{eq:dtheta_dm}, for $j=1,2,3$,
\begin{equation}
\mathbf U(\boldsymbol\beta_{j})=\sum_{i=1}^n\mathbf f_{j}(\mathbf x_i)\,\frac1{g_{j}'(m_{j1i})}
\sum_{k=1}^3[\nabla^2b_1(\boldsymbol\theta_i)]^{-1}_{kj}\,\bigl(y_{ki}-m_{k\delta_ii}\bigr),
\label{eq:score_general_link}
\end{equation}
and the complete score is
$\mathbf U(\boldsymbol\beta)=\bigl(\mathbf U(\boldsymbol\beta_{1})^\trp,\mathbf U(\boldsymbol\beta_{2})^\trp,\mathbf U(\boldsymbol\beta_{3})^\trp\bigr)^\trp$. For $\delta_i=0$, $m_{k0i}=\partial b_0/\partial\theta_{ki}$ is a function of $\boldsymbol\beta$ through
$\boldsymbol\theta_i=H_1(\boldsymbol m_{1i})$, so observations in both regions contribute to the
estimation of every $\boldsymbol\beta_{j}$.

\begin{remark} Under the log link, for $g_{j}=\log$, $1/g_{j}'(m_{j1i})=m_{j1i}$ and
\begin{equation}
\mathbf U(\boldsymbol\beta_{j})=\sum_{i=1}^n\mathbf f_{j}(\mathbf x_i)\,m_{j1i}
\sum_{k=1}^3[\nabla^2b_1(\boldsymbol\theta_i)]^{-1}_{kj}\,\bigl(y_{ki}-m_{k\delta_ii}\bigr).
\label{eq:score_log}
\end{equation}
The link requires $m_{j1i}>0$. Here $m_{11i}>0$ and $m_{31i}>0$ always, whereas
$m_{21i}=m_{31i}+A_{1i}$ with $A_{1i}<0$ by \eqref{eq:compat_identity}, so $m_{21i}>0$ means
$m_{31i}>-A_{1i}$. This is an additional restriction on the regression parameters, to be
imposed over the whole design space together with $\theta_{ji}<0$. A simpler log-link model puts the link on the rates,
$\lambda_{ji}=\exp\{\mathbf f_j(\mathbf x_i)^\trp\boldsymbol\beta_j\}$. Then $\theta_{ji}=-\lambda_{ji}$ and
$\partial\theta_{ki}/\partial\boldsymbol\beta_j=-\lambda_{ji}\mathbf f_j(\mathbf x_i)$ for $k=j$ and $\mathbf 0$ otherwise, so
\[
\mathbf U(\boldsymbol\beta_j)=-\sum_{i=1}^n\lambda_{ji}\,\mathbf f_j(\mathbf x_i)\,\bigl(y_{ji}-m_{j\delta_ii}\bigr),\qquad j=1,2,3,
\]
with no inversion of $H_r$, no anchor region, and $\lambda_{ji}>0$ automatically.

For the canonical link, with $\theta_{ji}=\mathbf f_j(\mathbf x_i)^\trp\boldsymbol\beta_j$ we have
$\partial\theta_{ki}/\partial\boldsymbol\beta_j=\mathbf f_j(\mathbf x_i)$ for $k=j$ and $\mathbf 0$ otherwise, so
$\mathbf U(\boldsymbol\beta_j)=\sum_{i=1}^n\mathbf f_j(\mathbf x_i)\,(y_{ji}-m_{j\delta_ii})$.
\end{remark}

  Under the usual regularity conditions, the Fisher information matrix $\mathcal I(\boldsymbol\beta)=\mathbb E\{\mathbf U(\boldsymbol\beta)\mathbf U(\boldsymbol\beta)^\trp\}
=-\mathbb E\{\partial^2\ell(\boldsymbol\beta)/\partial\boldsymbol\beta\,\partial\boldsymbol\beta^\trp\}$.
Differentiating the score and using $\mathbb E(\mathbf T_i-\boldsymbol m_{\delta_ii})=\mathbf 0$
(the $\pi_{ri}$-weighted gradients $\nabla\pi_{1i}+\nabla\pi_{0i}$ cancel) gives, since
$\mathbb E\{\nabla^2b_{\delta_i}(\boldsymbol\theta_i)\}=\sum_r\pi_{ri}\nabla^2b_r(\boldsymbol\theta_i)$,
\begin{equation}
\mathcal I(\boldsymbol\beta)=\sum_{i=1}^n\sum_{r=0}^1\pi_{ri}
\sum_{k=1}^3\sum_{\ell=1}^3
\frac{\partial^2b_r(\boldsymbol\theta_i)}{\partial\theta_{ki}\,\partial\theta_{\ell i}}\,
\frac{\partial\theta_{ki}}{\partial\boldsymbol\beta}\Bigl(\frac{\partial\theta_{\ell i}}{\partial\boldsymbol\beta}\Bigr)^{\!\trp},
\label{eq:fisher_block_expanded}
\end{equation}
where $\partial\theta_{ki}/\partial\boldsymbol\beta$ is the $p$-vector whose $\boldsymbol\beta_{j}$-block is
\eqref{eq:score_chain}, $j=1,2,3$. The Fisher information matrix \eqref{eq:fisher_block_expanded} is generally not block
diagonal. Equivalently, writing
$\mathcal I(\boldsymbol\beta)
=\{\mathcal I_{js}(\boldsymbol\beta)\}_{j,s=1}^{3}$,
the $(j,s)$ block is
\begin{equation}
\mathcal I_{js}(\boldsymbol\beta)
=
\sum_{i=1}^{n}
\frac{\mathbf f_{j}(x_i)\mathbf f_{s}(x_i)^{\trp}}
{g_j'(m_{j1i})\,g_s'(m_{s1i})}
\sum_{r=0}^{1}\pi_{ri}
\left[
\{\nabla^2 b_1(\boldsymbol\theta_i)\}^{-1}
\nabla^2 b_r(\boldsymbol\theta_i)
\{\nabla^2 b_1(\boldsymbol\theta_i)\}^{-1}
\right]_{js},
\qquad j,s=1,2,3.
\end{equation}
If all components share the same design vector, the outer product
reduces to $\mathbf f(x_i)\mathbf f(x_i)^\trp$. The entries of $\nabla^2b_1$ and
$\nabla^2b_0$ follow from $b_1,b_0$ in \eqref{eq:b1-theta}--\eqref{eq:b0-theta}
(symmetric matrices; upper triangle shown):
\begin{align*}
\tfrac{\partial^2b_1}{\partial\theta_{1i}^2}&=\tfrac1{\theta_i^2}+\tfrac1{\theta_{1i}^2}-\tfrac1{(\theta_{1i}+\theta_{2i})^2},&
\tfrac{\partial^2b_1}{\partial\theta_{1i}\partial\theta_{2i}}&=\tfrac1{\theta_i^2}-\tfrac1{(\theta_{1i}+\theta_{2i})^2},&
\tfrac{\partial^2b_1}{\partial\theta_{1i}\partial\theta_{3i}}&=\tfrac1{\theta_i^2},\\
\tfrac{\partial^2b_1}{\partial\theta_{2i}^2}&=\tfrac1{\theta_i^2}+\tfrac1{(\theta_{2i}+\theta_{3i})^2}-\tfrac1{(\theta_{1i}+\theta_{2i})^2},&
\tfrac{\partial^2b_1}{\partial\theta_{2i}\partial\theta_{3i}}&=\tfrac1{\theta_i^2}+\tfrac1{(\theta_{2i}+\theta_{3i})^2},&
\tfrac{\partial^2b_1}{\partial\theta_{3i}^2}&=\tfrac1{\theta_i^2}+\tfrac1{(\theta_{2i}+\theta_{3i})^2},
\end{align*}
\begin{align*}
\tfrac{\partial^2b_0}{\partial\theta_{1i}^2}&=\tfrac1{\theta_i^2}+\tfrac1{(\theta_{1i}+\theta_{3i})^2}-\tfrac1{(\theta_{1i}+\theta_{2i})^2},&
\tfrac{\partial^2b_0}{\partial\theta_{1i}\partial\theta_{2i}}&=\tfrac1{\theta_i^2}-\tfrac1{(\theta_{1i}+\theta_{2i})^2},&
\tfrac{\partial^2b_0}{\partial\theta_{1i}\partial\theta_{3i}}&=\tfrac1{\theta_i^2}+\tfrac1{(\theta_{1i}+\theta_{3i})^2},\\
\tfrac{\partial^2b_0}{\partial\theta_{2i}^2}&=\tfrac1{\theta_i^2}+\tfrac1{\theta_{2i}^2}-\tfrac1{(\theta_{1i}+\theta_{2i})^2},&
\tfrac{\partial^2b_0}{\partial\theta_{2i}\partial\theta_{3i}}&=\tfrac1{\theta_i^2},&
\tfrac{\partial^2b_0}{\partial\theta_{3i}^2}&=\tfrac1{\theta_i^2}+\tfrac1{(\theta_{1i}+\theta_{3i})^2}.
\end{align*}

An  maximum-likelihood estimate satisfies $\mathbf U(\widehat{\boldsymbol\beta})=\mathbf 0$. When
$\mathcal I(\boldsymbol\beta)$ is nonsingular, Fisher scoring gives
\begin{equation}
\boldsymbol\beta^{(t+1)}=\boldsymbol\beta^{(t)}+\mathcal I\{\boldsymbol\beta^{(t)}\}^{-1}\mathbf U\{\boldsymbol\beta^{(t)}\}.
\label{eq:fisher_scoring_complete}
\end{equation}
The step is shortened when necessary so that the updated parameters stay admissible, i.e.\
$\theta_{ji}<0$ over the design space, $\nabla^2b_1(\boldsymbol\theta_i)$ is invertible, and, for
the log link, $m_{21i}>0$.

\section{A GLM formulation for the conditional joint distributions}\label{sec-5}

In the full model developed in Section \ref{sec-4}, the regression coefficients usually do not have a direct interpretation as effects on the marginal
means. For this reason, we now consider the normalized conditional joint distribution on a chosen region. In this distribution, the cumulant
derivatives are exactly the conditional means. The link functions then connect these conditional means to the covariates, so the coefficients
can be interpreted as effects on the conditional means inside this region. These effects depend on the ordering of the responses ($Y_1<Y_2$
for $R_1$), so they are not effects on the marginal means. Also, this model uses only the observations that fall in the chosen region.

 In the following we develop a GLM for conditional joint densities defined in Lemma \ref{lem:cond-dens}. We focus on
 $P_{R_1}(\BS{y}_i; \BS{\lambda}_i)=\lambda_{23i}\lambda_i\,\exp\!\big[-\lambda_{1i} y_{1i}-\lambda_{23i}y_{2i}\big]$  on the region $R_1$. Denote  $\theta_{1i}=-\lambda_{1i}$, $\theta_{23i}=-\lambda_{23i}$, and $\theta_i=-\lambda_i$ so that $\theta_{1i}<0$, $\theta_{23i}<0$, $\theta_i<0$. Given $\BS{\theta}_i=(\theta_{1i}, \theta_{23i})^\trp$,  the conditional joint densities can be written in canonical exponential-family form
\begin{equation}
P_{R_1}(\BS{y}_i;\boldsymbol{\theta}_i)
=
\exp\!\left\{
\boldsymbol{\theta}_i^\trp \BS{y}_i-b(\boldsymbol{\theta}_i)
\right\},
\qquad (y_{1i},y_{2i})^\trp\in R_1
\end{equation}
where the cumulant function is given by $b(\boldsymbol{\theta}_i) =-\log(-\theta_i)-\log\!\bigl(-\theta_{23i}\bigr)$. The moments are then obtained as 
\[
\boldsymbol{\mu}_i
=\nabla b(\boldsymbol{\theta}_i)
=
\begin{pmatrix}
-\dfrac{1}{\theta_i}
\\[3mm]
-\dfrac{1}{\theta_{23i}}
-\dfrac{1}{\theta_i}
\end{pmatrix},\,\,
\mathbf V_i=\nabla^2 b(\boldsymbol{\theta}_i)
=
\begin{pmatrix}
\dfrac{1}{\theta_i^2}
&
\dfrac{1}{\theta_i^2}
\\[4mm]
\dfrac{1}{\theta_i^2}
&
\dfrac{1}{\theta_{i}^2}
+\dfrac{1}{\theta_{23i}^2}
\end{pmatrix}, 
\]
with
\[
\mathbf V_i^{-1}
=
\begin{pmatrix}
\theta_i^2+\theta_{23i}^2
&
-\theta_{23i}^2
\\
-\theta_{23i}^2
&
\theta_{23i}^2
\end{pmatrix}.
\]
From the formulas of the conditional means $\mu_{1i}$ and $\mu_{2i}$, it is straightforward to induce the natural parameters 
\begin{equation}
\theta_{1i}=-\frac1{\mu_{1i}}+\frac1{\mu_{2i}-\mu_{1i}},
\qquad
\theta_{23i}=-\frac1{\mu_{2i}-\mu_{1i}}.
\label{eq:R1_inverse_mean}
\end{equation}
Since $\theta_{1i}<0$, we obtain $\mu_{1i}>0$ and $\mu_{2i}>2\mu_{1i}$. This represents parameter restrictions. 
For each component $j=1,2$,  the link functions are defined by $\eta_{j}=g_j(\mu_{j})$, where  $\eta_{j}=\mathbf f_j(\mathbf x)^{\trp}\boldsymbol{\beta}_j$, 
$\mathbf f_j(\mathbf x)=\bigl(1,f_{j1}(\mathbf x),\ldots,f_{jk_j}(\mathbf x)\bigr)^{\trp}$ and  $\boldsymbol{\beta}_j=\bigl(\beta_{j0},\ldots,\beta_{jk_j}\bigr)^{\trp}$. The two components may have different links, covariates and coefficients. The complete $p$-dimensional coefficient vector is $\BS{\beta}=(\BS{\beta}_1^\trp,\BS{\beta}_2^\trp)^\trp$ where $p=\sum_{j=1}^{2}(k_j+1)=k_1+k_2+2$.

\subsection{Scores and Fisher Information}

The log-likelihood contribution of observation $i$ is $\ell_i\{\boldsymbol\theta_i(\boldsymbol\beta)\}=\theta_{1i}(\boldsymbol\beta)y_{1i}+\theta_{23,i}(\boldsymbol\beta)y_{2i}-b\{\boldsymbol\theta_i(\boldsymbol\beta)\}$ and  $\ell(\boldsymbol\beta)=\sum_{i=1}^n\ell_i\{\boldsymbol\theta_i(\boldsymbol\beta)\}$.  Since $\boldsymbol\mu_i=\nabla b(\boldsymbol\theta_i)$,
\begin{equation}
\frac{\partial\ell_i}{\partial\boldsymbol\theta_i}=\boldsymbol y_i-\boldsymbol\mu_i,\qquad
\frac{\partial\boldsymbol\mu_i}{\partial\boldsymbol\theta_i^\trp}=\mathbf V_i .
\label{eq:c_dl_dtheta}
\end{equation}
Hence $\partial\boldsymbol\theta_i/\partial\boldsymbol\mu_i^\trp=\mathbf V_i^{-1}$; in terms of the means, 
\begin{equation}
\mathbf V_i^{-1}=
\begin{pmatrix}
\dfrac1{\mu_{1i}^2}+\dfrac1{(\mu_{2i}-\mu_{1i})^2}&-\dfrac1{(\mu_{2i}-\mu_{1i})^2}\\[3mm]
-\dfrac1{(\mu_{2i}-\mu_{1i})^2}&\dfrac1{(\mu_{2i}-\mu_{1i})^2}
\end{pmatrix}.
\label{eq:c_dtheta_dmu}
\end{equation}
Each component of $\boldsymbol\theta_i$ depends on both means, while $\eta_{\ell i}$ depends only on
$\boldsymbol\beta_{\ell}$. Hence, for $j=1,2$,
\begin{equation}
\frac{\partial\boldsymbol\theta_i}{\partial\boldsymbol\beta_{j}^\trp}
=\sum_{\ell=1}^2\frac{\partial\boldsymbol\theta_i}{\partial\mu_{\ell i}}
\frac{\partial\mu_{\ell i}}{\partial\eta_{\ell i}}
\frac{\partial\eta_{\ell i}}{\partial\boldsymbol\beta_{j}^\trp}
=\frac1{g_{j}'(\mu_{ji})}\,[\mathbf V_i^{-1}]_{\cdot j}\,\mathbf f_{j}(\mathbf x_i)^\trp,
\label{eq:c_score_chain}
\end{equation}
because only the term $\ell=j$ survives, where $[\mathbf V_i^{-1}]_{\cdot j}$ denotes the $j$th column of
$\mathbf V_i^{-1}$. The factor $\partial\theta_{1i}/\partial\mu_{1i}=1/\mu_{1i}^2+1/(\mu_{2i}-\mu_{1i})^2$ is
\emph{not} $1/\mu_{1i}^2$. Combining \eqref{eq:c_dl_dtheta}--\eqref{eq:c_score_chain}, for $j=1,2$,
\begin{equation}
\mathbf U(\boldsymbol\beta_{j})=\sum_{i=1}^n\mathbf f_{j}(\mathbf x_i)\,\frac1{g_{j}'(\mu_{ji})}
\sum_{k=1}^2[\mathbf V_i^{-1}]_{kj}\,(y_{ki}-\mu_{ki}),
\label{eq:c_score_general}
\end{equation}
and the complete score is $\mathbf U(\boldsymbol\beta)=\bigl(\mathbf U(\boldsymbol\beta_1)^\trp,\mathbf U(\boldsymbol\beta_2)^\trp\bigr)^\trp$.

\begin{remark}
Under the log link, $g_j=\log$ and $1/g_j'(\mu_{ji})=\mu_{ji}$, so by \eqref{eq:c_dtheta_dmu}
\begin{align}
\mathbf U(\boldsymbol\beta_1)&=\sum_{i=1}^n\mathbf f_1(\mathbf x_i)\Bigl[\frac{y_{1i}-\mu_{1i}}{\mu_{1i}}
+\frac{\mu_{1i}\{(y_{1i}-\mu_{1i})-(y_{2i}-\mu_{2i})\}}{(\mu_{2i}-\mu_{1i})^2}\Bigr],\label{eq:c_score_log1}\\
\mathbf U(\boldsymbol\beta_2)&=\sum_{i=1}^n\mathbf f_2(\mathbf x_i)\,
\frac{\mu_{2i}\{(y_{2i}-\mu_{2i})-(y_{1i}-\mu_{1i})\}}{(\mu_{2i}-\mu_{1i})^2}.\label{eq:c_score_log2}
\end{align}
Thus each score involves both response components, and $\eta_{2i}-\eta_{1i}>\log2$ must hold over the whole design
space. A simpler log-link model puts the link on the rates, $\lambda_{1i}=\exp\{\mathbf f_1(\mathbf x_i)^\trp\boldsymbol\beta_1\}$ and
$\lambda_{23,i}=\exp\{\mathbf f_2(\mathbf x_i)^\trp\boldsymbol\beta_2\}$. Then $\theta_{1i}=-\lambda_{1i}$,
$\theta_{23,i}=-\lambda_{23,i}$, and $\boldsymbol\theta_i$ depends on $\boldsymbol\beta_1$ only through $\theta_{1i}$ and on
$\boldsymbol\beta_2$ only through $\theta_{23,i}$, with derivatives $-\lambda_{1i}\mathbf f_1(\mathbf x_i)$ and
$-\lambda_{23,i}\mathbf f_2(\mathbf x_i)$, so
\[
\mathbf U(\boldsymbol\beta_1)=-\sum_{i=1}^n\lambda_{1i}\,\mathbf f_1(\mathbf x_i)\,(y_{1i}-\mu_{1i}),\qquad
\mathbf U(\boldsymbol\beta_2)=-\sum_{i=1}^n\lambda_{23,i}\,\mathbf f_2(\mathbf x_i)\,(y_{2i}-\mu_{2i}),
\]
with no cross terms from the Jacobian and only $\lambda_{1i},\lambda_{23,i}>0$ to be imposed.

For the canonical link, $\partial\theta_{1i}/\partial\boldsymbol\beta_1=\mathbf f_1(\mathbf x_i)$ and
$\partial\theta_{23,i}/\partial\boldsymbol\beta_2=\mathbf f_2(\mathbf x_i)$, with the other derivatives zero, so
$\mathbf U(\boldsymbol\beta_j)=\sum_{i=1}^n\mathbf f_j(\mathbf x_i)(y_{ji}-\mu_{ji})$, $j=1,2$.
\end{remark}

Under the usual regularity conditions,
$\mathcal I(\boldsymbol\beta)=\mathbb E\{\mathbf U(\boldsymbol\beta)\mathbf U(\boldsymbol\beta)^\trp\}
=-\mathbb E\{\partial^2\ell(\boldsymbol\beta)/\partial\boldsymbol\beta\,\partial\boldsymbol\beta^\trp\}$.
Differentiating the score and using $\mathbb E(\boldsymbol y_i-\boldsymbol\mu_i)=\mathbf 0$ and
$\nabla^2b(\boldsymbol\theta_i)=\mathbf V_i$ gives
\begin{equation}
\mathcal I(\boldsymbol\beta)=\sum_{i=1}^n\Bigl(\frac{\partial\boldsymbol\theta_i}{\partial\boldsymbol\beta^\trp}\Bigr)^{\!\trp}
\mathbf V_i\Bigl(\frac{\partial\boldsymbol\theta_i}{\partial\boldsymbol\beta^\trp}\Bigr)
=\sum_{i=1}^n\mathbf F_i^\trp\mathbf D_i\mathbf V_i^{-1}\mathbf D_i\mathbf F_i,
\label{eq:conditional_information}
\end{equation}
where $\mathbf D_i=\operatorname{diag}\{1/g_1'(\mu_{1i}),1/g_2'(\mu_{2i})\}$ and
$\mathbf F_i=\operatorname{blockdiag}\{\mathbf f_1(\mathbf x_i)^\trp,\mathbf f_2(\mathbf x_i)^\trp\}$.
Equivalently, writing $\mathcal I(\boldsymbol\beta)=\{\mathcal I_{js}(\boldsymbol\beta)\}_{j,s=1}^2$, the $(j,s)$ block is
\begin{equation}
\mathcal I_{js}(\boldsymbol\beta)=\sum_{i=1}^n
\frac{\mathbf f_j(\mathbf x_i)\mathbf f_s(\mathbf x_i)^\trp}{g_j'(\mu_{ji})\,g_s'(\mu_{si})}\,[\mathbf V_i^{-1}]_{js},
\qquad j,s=1,2 .
\label{eq:c_fisher_blocks}
\end{equation}
This matrix is generally not block diagonal, because $[\mathbf V_i^{-1}]_{12}=-1/(\mu_{2i}-\mu_{1i})^2\neq0$.
If both components share the same design vector, the outer product reduces to $\mathbf f(\mathbf x_i)\mathbf f(\mathbf x_i)^\trp$.

\begin{remark}
Under the canonical link, $\boldsymbol{\eta}_i=\boldsymbol{\theta}_i$, we have $\partial\boldsymbol{\mu}_i/\partial\boldsymbol{\eta}_i^{\trp}
=\mathbf V_i$, where
\[
\mathbf V_i=
\begin{pmatrix}
\mu_{1i}^2 & \mu_{1i}^2\\
\mu_{1i}^2 & \mu_{1i}^2+(\mu_{2i}-\mu_{1i})^2
\end{pmatrix}.
\]
The component scores are $\mathbf U_j(\boldsymbol\beta)=\sum_{i=1}^{n}\mathbf f_j(\mathbf x_i)(y_{ji}-\mu_{ji}),  j=1,2$, and the Fisher information is $
\mathcal I(\boldsymbol\beta)=\sum_{i=1}^{n}\mathbf F_i^\trp\mathbf V_i\mathbf F_i$. Thus, its blocks are $\mathcal I_{js}(\boldsymbol\beta)
=\sum_{i=1}^{n}[\mathbf V_i]_{js}\mathbf f_j(\mathbf x_i)\mathbf f_s(\mathbf x_i)^\trp, j,s=1,2$.  Each component score contains its corresponding response residual,
although the regression parameters are jointly estimated.
The restrictions $\eta_{1i}<0$ and $\eta_{2i}<0$
must hold throughout the design space.

Under the log link on the conditional means,
$\eta_{ji}=\log\mu_{ji}$, we have
$\mathbf D_i=\operatorname{diag}(\mu_{1i},\mu_{2i})$.
Each component score then involves both response residuals.
By \eqref{eq:c_fisher_blocks}, the information blocks are
\begin{align}
\mathcal I_{11}(\boldsymbol{\beta})
&=
\sum_{i=1}^{n}
\left\{
1+\frac{\mu_{1i}^2}{(\mu_{2i}-\mu_{1i})^2}
\right\}
\mathbf f_1(\mathbf x_i)\mathbf f_1(\mathbf x_i)^\trp,
\\
\mathcal I_{12}(\boldsymbol{\beta})
&=
-\sum_{i=1}^{n}
\frac{\mu_{1i}\mu_{2i}}{(\mu_{2i}-\mu_{1i})^2}
\mathbf f_1(\mathbf x_i)\mathbf f_2(\mathbf x_i)^\trp,
\\
\mathcal I_{22}(\boldsymbol{\beta})
&=
\sum_{i=1}^{n}
\frac{\mu_{2i}^2}{(\mu_{2i}-\mu_{1i})^2}
\mathbf f_2(\mathbf x_i)\mathbf f_2(\mathbf x_i)^\trp,
\end{align}
with
$\mathcal I_{21}(\boldsymbol{\beta})
=\mathcal I_{12}(\boldsymbol{\beta})^\trp$.
The restriction $\eta_{2i}-\eta_{1i}>\log 2$
must hold throughout the design space.

For the alternative log link on the rates,
$\eta_{1i}=\log\lambda_{1i}$ and
$\eta_{2i}=\log\lambda_{23,i}$, define
$\mathbf L_i=\operatorname{diag}(\lambda_{1i},\lambda_{23,i})$.
The Fisher information is
\[
\mathcal I(\boldsymbol\beta)
=
\sum_{i=1}^{n}
\mathbf F_i^\trp\mathbf L_i\mathbf V_i\mathbf L_i\mathbf F_i.
\]
The inverse links ensure that both rates are positive.
\end{remark}

An  maximum-likelihood estimate satisfies $\mathbf U(\widehat{\boldsymbol\beta})=\mathbf 0$. When
$\mathcal I(\boldsymbol\beta)$ is nonsingular, Fisher scoring gives
\begin{equation}
\boldsymbol\beta^{(t+1)}=\boldsymbol\beta^{(t)}+\mathcal I\{\boldsymbol\beta^{(t)}\}^{-1}\mathbf U\{\boldsymbol\beta^{(t)}\}.
\label{eq:c_fisher_scoring}
\end{equation}
The step is shortened when necessary so that the updated parameters stay admissible, i.e.\ $\mu_{1i}>0$ and
$\mu_{2i}>2\mu_{1i}$ for the mean links, or $\theta_{1i},\theta_{23,i}<0$ for the canonical link, over the design space.

\section{Simulation study}\label{sec-6}

We conducted a simulation study to examine estimation under
the full and conditional GLM formulations in
Sections~\ref{sec-4} and~\ref{sec-5}. Separate samples were
generated for the full model and the conditional model on
$R_1$. For each sample size $n\in\{200,500\}$, we generated
$B=1000$ independent samples. Each sample contained independent
observations with a fixed binary covariate $x\in\{0,1\}$,
with $n/2$ observations at each covariate level.
For each component, the linear predictor was
$\eta_j(x)=\beta_{j0}+\beta_{j1}x$.
Within each model, the canonical and log links were fitted
to the same samples. Here, $n$ is the  sample size for
the full model and the conditional sample size for the
conditional model.

Throughout this section, $\mu_j^c(x)$ denotes the conditional
mean on $R_1$, while $\mu_j(x)$ denotes the unconditional
marginal mean of the full model.

The generating rate vectors were selected as $\boldsymbol\lambda(0)=(3,1,1.5)^\trp$ at the covariate value $x=0$ and
$\boldsymbol\lambda(1)=(4,1,2)^\trp$ at the covariate value $x=1$.  Hence, $\pi_1(0)=0.75$ and $\pi_1(1)=0.80$.  

For the full model with a log link, the linked quantities were given as $m_{j1}(x)=\partial b_1\{\boldsymbol\theta(x)\}/\partial\theta_j$, where
 $\boldsymbol\theta(x)=-\boldsymbol\lambda(x)$, and were positive at both covariate levels. For the conditional model under the log link, the linked
quantities were $\mu_j^c(x)$, which satisfied $\mu_2^c(x)>2\mu_1^c(x)>0$. This restriction follows from $\lambda_1(x)>0$.  Thus, the corresponding linear predictors were specified as \(\eta_j(x)=\log m_{j1}(x)\) and \(\eta_j(x)=\log\mu_j^c(x)\), respectively.

For each model and link function, the true regression coefficients were determined by $\beta_{j0}=\eta_j(0)$ and
$\beta_{j1}=\eta_j(1)-\eta_j(0)$.

For the canonical link, the linear predictors were the three natural parameters $\theta_1,\theta_{2},\theta_{3}$
in the full model while $\theta_1,\theta_{23}$ in the conditional
model.  The true regression parameters $\BS{\beta}_j$, reported in
Table~\ref{tab:true_regression_parameters}, were obtained
from $\beta_{j0}=\eta_j(0)$ and $\beta_{j1}=\eta_j(1)-\eta_j(0)$. Specifically, for the full model under the log link,  $\beta_{j0}=\log m_{j1}(0)$, $\beta_{j1}=\log\left\{m_{j1}(1)/m_{j1}(0)\right\}, j=1,2,3$. Thus, $\boldsymbol m_1(0)
=\left(\frac{35}{132},\frac{73}{220},\frac{32}{55}\right)^\trp$ and $\boldsymbol m_1(1)=\left(\frac{27}{140},\frac{29}{105},\frac{10}{21}\right)^\trp$. For example, the first component has $\beta_{10}=\log(35/132)\approx\ -1.327454$ and $\beta_{11}=\log\{(27/140)/(35/132)\}\approx\ -0.318352$. For the conditional model on $R_1$,  $\beta_{j0}=\log\mu_j^c(0)$, $\beta_{j1}=\log\left\{\mu_j^c(1)/\mu_j^c(0)\right\}, j=1,2$. Thus $\boldsymbol\mu^c(0) =\left(\frac{2}{11},\frac{32}{55}\right)^\trp$, $\boldsymbol\mu^c(1)=\left(\frac{1}{7},\frac{10}{21}\right)^\trp$. Here, for its first component, $\beta_{10}=\log(2/11)\approx\ -1.704748$ and $\beta_{11}=\log\{(1/7)/(2/11)\}\approx\ -0.241162$.

\begin{table}[htbp]
\centering
\caption{True regression parameters}
\label{tab:true_regression_parameters}
\small
\begin{tabular}{llrrrr}
\hline
& & \multicolumn{2}{c}{Canonical link}
  & \multicolumn{2}{c}{Log link}\\
\cline{3-4}\cline{5-6}
Model & Component $j$
& $\beta_{j0}$ & $\beta_{j1}$
& $\beta_{j0}$ & $\beta_{j1}$\\
\hline
Full & 1
& $-3.000000$ & $-1.000000$
& $-1.327454$ & $-0.318352$\\
& 2
& $-1.000000$ & $0.000000$
& $-1.103168$ & $-0.183496$\\
& 3
& $-1.500000$ & $-0.500000$
& $-0.541597$ & $-0.200340$\\
\hline
Conditional & 1
& $-3.000000$ & $-1.000000$
& $-1.704748$ & $-0.241162$\\
& 2
& $-2.500000$ & $-0.500000$
& $-0.541597$ & $-0.200340$\\
\hline
\end{tabular}
\end{table}

At each covariate level $x$, we independently generated
$\delta\sim\operatorname{Bernoulli}\{\pi_1(x)\}$ and
$Z\sim\operatorname{Exp}\{\lambda_1(x)+\lambda_2(x)+\lambda_3(x)\}$.
Given $\delta$, we generated $W$ independently of $Z$, with
rate $\lambda_2(x)+\lambda_3(x)$ when $\delta=1$ and
$\lambda_1(x)+\lambda_3(x)$ when $\delta=0$.
We then set $(Y_1,Y_2)=(Z,Z+W)$ when $\delta=1$ and
$(Y_1,Y_2)=(Z+W,Z)$ when $\delta=0$.
For the conditional model on $R_1$, we generated independent
$Z$ and $W$ with rates
$\lambda_1(x)+\lambda_2(x)+\lambda_3(x)$ and
$\lambda_2(x)+\lambda_3(x)$, respectively, and set
$(Y_1,Y_2)=(Z,Z+W)$.

For the binary covariate, the joint response model was fitted
within each covariate group, $x=0$ and $x=1$. The fitted
predictors were then transformed into regression intercepts
and slopes, i.e., $\widehat\beta_{j0}=\widehat\eta_j(0)$ and
$\widehat\beta_{j1}=\widehat\eta_j(1)-\widehat\eta_j(0)$. For each regression coefficient $\beta_{ja}$, $a\in\{0,1\}$,
we assessed empirical bias, empirical standard deviation (SD), average estimated standard error (SE), and coverage of nominal
$95\%$ Wald confidence intervals. For replication $b$, let $\widehat\beta_{ja}^{(b)}$ denote
the coefficient estimate. Its estimated standard error,
$\widehat{\mathrm{SE}}_{ja}^{(b)}$, was the square root
of the corresponding diagonal entry of
$\mathcal I(\widehat{\boldsymbol\beta}^{(b)})^{-1}$.
For each setting, let $B_{\mathrm{int}}$ denote the number
of retained admissible interior fits, indexed by
$b=1,\ldots,B_{\mathrm{int}}$. Let $\overline{\widehat\beta}_{ja}
=B_{\mathrm{int}}^{-1}
\sum_{b=1}^{B_{\mathrm{int}}}\widehat\beta_{ja}^{(b)}$. The performance measures were
\begin{align}
\operatorname{Bias}(\widehat\beta_{ja})
&=
\overline{\widehat\beta}_{ja}-\beta_{ja},
\\
\operatorname{SD}(\widehat\beta_{ja})
&=
\left\{
\frac{1}{B_{\mathrm{int}}-1}
\sum_{b=1}^{B_{\mathrm{int}}}
\left(
\widehat\beta_{ja}^{(b)}
-\overline{\widehat\beta}_{ja}
\right)^2
\right\}^{1/2},
\\
\operatorname{Average\,SE}(\widehat\beta_{ja})
&=
\frac{1}{B_{\mathrm{int}}}
\sum_{b=1}^{B_{\mathrm{int}}}
\widehat{\mathrm{SE}}_{ja}^{(b)},
\\
\operatorname{Coverage}(\widehat\beta_{ja})
&=
\frac{100}{B_{\mathrm{int}}}
\sum_{b=1}^{B_{\mathrm{int}}}
1\left\{
\left|\widehat\beta_{ja}^{(b)}-\beta_{ja}\right|
\leq
1.96\,\widehat{\mathrm{SE}}_{ja}^{(b)}
\right\}.
\end{align}
For the full model at $n=200$, three replications gave boundary
solutions under the canonical link and were excluded. Under
the log link, five replications were excluded; three gave
boundary solutions, and two did not give valid 
estimates within the log-link domain.  Therefore, the results were calculated using
$997$ replications for the canonical link and $995$ for
the log link. For all other settings, all $1000$ replications were kept. All performance measures were calculated using the retained
replications.

Tables~\ref{tab:simulation_full} and
\ref{tab:simulation_conditional} report the results for each
regression coefficient. Empirical SD decreased as the sample
size increased. For the conditional model, average SE was
close to empirical SD, and coverage ranged from 94.4\% to
95.7\%. Average SE and empirical SD were also generally
close for the full model under the canonical link.

For the full model under the log link, the second component
showed larger finite-sample effects. The bias of its intercept
changed from $-0.0746$ at $n=200$ to $-0.0235$ at $n=500$.
The difference between average SE and empirical SD also
became smaller. Coverage for the slope remained above the
nominal 95\% level, at 99.4\% and 98.1\%, respectively.
Thus, some Wald intervals were conservative in these settings.

\begin{table}[H]
\centering
\caption{Estimation results under the full GLM}
\label{tab:simulation_full}
\small
\begin{tabular}{rlrrrrr}
\hline
$n$ & Parameter & True value & Bias
& Empirical SD & Average SE & Coverage (\%)\\
\hline
\multicolumn{7}{l}{\textit{Canonical link}}\\
200 & $\beta_{10}$ & $-3.0000$ & $-0.0273$ & 0.6153 & 0.6059 & 94.48\\
    & $\beta_{11}$ & $-1.0000$ & $-0.0150$ & 0.9789 & 0.9683 & 94.18\\
    & $\beta_{20}$ & $-1.0000$ & $-0.0157$ & 0.3192 & 0.3161 & 93.68\\
    & $\beta_{21}$ & $ 0.0000$ & $-0.0159$ & 0.4686 & 0.4607 & 96.69\\
    & $\beta_{30}$ & $-1.5000$ & $-0.0170$ & 0.4852 & 0.4950 & 95.79\\
    & $\beta_{31}$ & $-0.5000$ & $ 0.0028$ & 0.7288 & 0.7321 & 94.98\\[1mm]
500 & $\beta_{10}$ & $-3.0000$ & $-0.0098$ & 0.3924 & 0.3797 & 93.50\\
    & $\beta_{11}$ & $-1.0000$ & $ 0.0040$ & 0.6170 & 0.6053 & 95.00\\
    & $\beta_{20}$ & $-1.0000$ & $-0.0123$ & 0.1976 & 0.1987 & 94.00\\
    & $\beta_{21}$ & $ 0.0000$ & $ 0.0001$ & 0.2819 & 0.2853 & 96.80\\
    & $\beta_{30}$ & $-1.5000$ & $-0.0001$ & 0.3172 & 0.3101 & 94.80\\
    & $\beta_{31}$ & $-0.5000$ & $-0.0034$ & 0.4656 & 0.4551 & 94.90\\
\hline
\multicolumn{7}{l}{\textit{Log link}}\\
200 & $\beta_{10}$ & $-1.3275$ & $-0.0018$ & 0.1370 & 0.1333 & 94.97\\
    & $\beta_{11}$ & $-0.3184$ & $-0.0025$ & 0.1803 & 0.1828 & 95.88\\
    & $\beta_{20}$ & $-1.1032$ & $-0.0746$ & 0.3598 & 0.4064 & 96.68\\
    & $\beta_{21}$ & $-0.1835$ & $ 0.0177$ & 0.4491 & 0.5275 & 99.40\\
    & $\beta_{30}$ & $-0.5416$ & $-0.0039$ & 0.0833 & 0.0831 & 95.18\\
    & $\beta_{31}$ & $-0.2003$ & $-0.0023$ & 0.1135 & 0.1172 & 95.28\\[1mm]
500 & $\beta_{10}$ & $-1.3275$ & $ 0.0013$ & 0.0865 & 0.0828 & 94.50\\
    & $\beta_{11}$ & $-0.3184$ & $ 0.0002$ & 0.1147 & 0.1137 & 94.70\\
    & $\beta_{20}$ & $-1.1032$ & $-0.0235$ & 0.1713 & 0.1605 & 94.60\\
    & $\beta_{21}$ & $-0.1835$ & $ 0.0045$ & 0.2280 & 0.2191 & 98.10\\
    & $\beta_{30}$ & $-0.5416$ & $-0.0011$ & 0.0534 & 0.0526 & 94.60\\
    & $\beta_{31}$ & $-0.2003$ & $-0.0004$ & 0.0733 & 0.0741 & 95.40\\
\hline
\end{tabular}
\medskip
\end{table}

\begin{table}[H]
\centering
\caption{Estimation results under the conditional GLM on $R_1$.}
\label{tab:simulation_conditional}
\small
\begin{tabular}{rlrrrrr}
\hline
$n$ & Parameter & True value & Bias
& Empirical SD & Average SE & Coverage (\%)\\
\hline
\multicolumn{7}{l}{\textit{Canonical link}}\\
200 & $\beta_{10}$ & $-3.0000$ & $-0.0038$ & 0.6170 & 0.6081 & 95.00\\
    & $\beta_{11}$ & $-1.0000$ & $-0.0354$ & 0.9896 & 0.9829 & 95.10\\
    & $\beta_{20}$ & $-2.5000$ & $-0.0201$ & 0.2547 & 0.2520 & 95.70\\
    & $\beta_{21}$ & $-0.5000$ & $-0.0103$ & 0.4078 & 0.3951 & 94.40\\[1mm]
500 & $\beta_{10}$ & $-3.0000$ & $-0.0182$ & 0.3788 & 0.3835 & 95.40\\
    & $\beta_{11}$ & $-1.0000$ & $-0.0149$ & 0.6097 & 0.6187 & 94.50\\
    & $\beta_{20}$ & $-2.5000$ & $-0.0014$ & 0.1574 & 0.1582 & 94.60\\
    & $\beta_{21}$ & $-0.5000$ & $-0.0126$ & 0.2446 & 0.2479 & 95.40\\
\hline
\multicolumn{7}{l}{\textit{Log link}}\\
200 & $\beta_{10}$ & $-1.7047$ & $ 0.0006$ & 0.0996 & 0.1000 & 95.40\\
    & $\beta_{11}$ & $-0.2412$ & $-0.0057$ & 0.1401 & 0.1414 & 95.20\\
    & $\beta_{20}$ & $-0.5416$ & $ 0.0004$ & 0.0742 & 0.0756 & 95.20\\
    & $\beta_{21}$ & $-0.2003$ & $-0.0032$ & 0.1076 & 0.1073 & 94.70\\[1mm]
500 & $\beta_{10}$ & $-1.7047$ & $-0.0015$ & 0.0635 & 0.0632 & 95.00\\
    & $\beta_{11}$ & $-0.2412$ & $-0.0031$ & 0.0879 & 0.0894 & 95.40\\
    & $\beta_{20}$ & $-0.5416$ & $ 0.0013$ & 0.0483 & 0.0478 & 94.40\\
    & $\beta_{21}$ & $-0.2003$ & $-0.0038$ & 0.0671 & 0.0679 & 95.20\\
\hline
\end{tabular}
\medskip
\end{table}

We also evaluated estimation of the conditional means
$\mu_j^c(x)$, the full-model cumulant derivatives
$m_{jr}(x)$, $j=1,2$, $r=0,1$, and the unconditional
marginal means
$\mu_j(x)=\pi_1(x)m_{j1}(x)+\pi_0(x)m_{j0}(x)$.
All estimates were calculated from the fitted
regression parameters in each retained replication, with
\[
\widehat \mu_j^{(b)}(x)
=\widehat\pi_1^{(b)}(x)\widehat m_{j1}^{(b)}(x)
+\widehat\pi_0^{(b)}(x)\widehat m_{j0}^{(b)}(x).
\]

For each target $q(x)$, let
$\overline{\widehat q}(x)
=B_{\rm int}^{-1}\sum_{b=1}^{B_{\rm int}}\widehat q^{(b)}(x)$
denote the average estimate over the retained replications.
The empirical bias and RMSE were calculated as
\begin{align}
\operatorname{Bias}\{\widehat q(x)\}
&=
\overline{\widehat q}(x)-q(x),\\
\operatorname{RMSE}\{\widehat q(x)\}
&=
\left\{
\frac{1}{B_{\rm int}}
\sum_{b=1}^{B_{\rm int}}
\left(\widehat q^{(b)}(x)-q(x)\right)^2
\right\}^{1/2}.
\end{align}

Tables~\ref{tab:additional_conditional}--
\ref{tab:additional_marginal} report the true values,
average estimates, empirical biases, and RMSEs.
The column ``Mean'' gives the average estimate over
the retained replications. The conditional and unconditional response means had absolute empirical
biases below $0.002$ in all settings. Their RMSEs decreased with sample
size. For example, under the conditional model with canonical link, the RMSE for
$\mu_2^c(0)$ decreased from $0.0430$ to $0.0281$. Under the full model with canonical
link, the RMSE for $\mu_2(0)$ decreased from $0.0431$ to $0.0269$.
Some cumulant derivatives were less accurately recovered.
In particular, for $m_{20}(1)$ under the full model with canonical link, bias
decreased from $0.0816$ to $0.0258$ and RMSE from $0.3734$ to $0.1943$.
RMSE decreased for every reported derivative under both links.
Thus, accurate recovery of the marginal means did not imply equally
accurate recovery of each region-specific cumulant derivative.

The two links gave identical conditional-mean summaries to the reported
precision, as expected for this saturated binary-covariate design.
The full-model summaries were also nearly identical, with small
differences at $n=200$ reflecting the different sets of retained fits.
Each estimator was assessed against its own population target;
these results do not compare efficiency between the full and
conditional experiments.

\begin{table}[H]
\centering\small
\caption{Estimated conditional means on $R_1$}
\label{tab:additional_conditional}
\begin{tabular}{rrlrrrr}
\hline
$n$ & $x$ & Target & True & Mean & Bias & RMSE \\
\hline
\multicolumn{7}{l}{\textit{Canonical link}} \\
\hline
200 & 0 & $\mu_1^c$ & 0.1818 & 0.1828 & 0.0010 & 0.0183 \\
200 & 0 & $\mu_2^c$ & 0.5818 & 0.5836 & 0.0018 & 0.0430 \\
200 & 1 & $\mu_1^c$ & 0.1429 & 0.1428 & 0.0000 & 0.0140 \\
200 & 1 & $\mu_2^c$ & 0.4762 & 0.4763 & 0.0001 & 0.0362 \\
500 & 0 & $\mu_1^c$ & 0.1818 & 0.1819 & 0.0001 & 0.0116 \\
500 & 0 & $\mu_2^c$ & 0.5818 & 0.5833 & 0.0014 & 0.0281 \\
500 & 1 & $\mu_1^c$ & 0.1429 & 0.1425 & -0.0004 & 0.0091 \\
500 & 1 & $\mu_2^c$ & 0.4762 & 0.4756 & -0.0006 & 0.0227 \\
\hline
\multicolumn{7}{l}{\textit{Log link}} \\
\hline
200 & 0 & $\mu_1^c$ & 0.1818 & 0.1828 & 0.0010 & 0.0183 \\
200 & 0 & $\mu_2^c$ & 0.5818 & 0.5836 & 0.0018 & 0.0430 \\
200 & 1 & $\mu_1^c$ & 0.1429 & 0.1428 & 0.0000 & 0.0140 \\
200 & 1 & $\mu_2^c$ & 0.4762 & 0.4763 & 0.0001 & 0.0362 \\
500 & 0 & $\mu_1^c$ & 0.1818 & 0.1819 & 0.0001 & 0.0116 \\
500 & 0 & $\mu_2^c$ & 0.5818 & 0.5833 & 0.0014 & 0.0281 \\
500 & 1 & $\mu_1^c$ & 0.1429 & 0.1425 & -0.0004 & 0.0091 \\
500 & 1 & $\mu_2^c$ & 0.4762 & 0.4756 & -0.0006 & 0.0227 \\
\hline
\end{tabular}
\end{table}

\begin{table}[H]
\centering\small
\caption{Estimated cumulant derivatives under the full model}
\label{tab:additional_derivatives}
\begin{tabular}{rrlrrrr}
\hline
$n$ & $x$ & Target & True & Mean & Bias & RMSE \\
\hline
\multicolumn{7}{l}{\textit{Canonical link}} \\
\hline
200 & 0 & $m_{11}$ & 0.2652 & 0.2673 & 0.0022 & 0.0376 \\
200 & 0 & $m_{21}$ & 0.3318 & 0.3221 & -0.0097 & 0.0827 \\
200 & 0 & $m_{10}$ & 0.1540 & 0.1439 & -0.0101 & 0.0466 \\
200 & 0 & $m_{20}$ & 0.9318 & 1.0069 & 0.0751 & 0.3248 \\
200 & 1 & $m_{11}$ & 0.1929 & 0.1936 & 0.0007 & 0.0242 \\
200 & 1 & $m_{21}$ & 0.2762 & 0.2690 & -0.0071 & 0.0636 \\
200 & 1 & $m_{10}$ & 0.1095 & 0.1030 & -0.0065 & 0.0321 \\
200 & 1 & $m_{20}$ & 0.9429 & 1.0244 & 0.0816 & 0.3734 \\
500 & 0 & $m_{11}$ & 0.2652 & 0.2665 & 0.0013 & 0.0234 \\
500 & 0 & $m_{21}$ & 0.3318 & 0.3286 & -0.0032 & 0.0523 \\
500 & 0 & $m_{10}$ & 0.1540 & 0.1508 & -0.0032 & 0.0274 \\
500 & 0 & $m_{20}$ & 0.9318 & 0.9549 & 0.0231 & 0.1851 \\
500 & 1 & $m_{11}$ & 0.1929 & 0.1937 & 0.0009 & 0.0153 \\
500 & 1 & $m_{21}$ & 0.2762 & 0.2737 & -0.0025 & 0.0374 \\
500 & 1 & $m_{10}$ & 0.1095 & 0.1075 & -0.0021 & 0.0175 \\
500 & 1 & $m_{20}$ & 0.9429 & 0.9686 & 0.0258 & 0.1943 \\
\hline
\multicolumn{7}{l}{\textit{Log link}} \\
\hline
200 & 0 & $m_{11}$ & 0.2652 & 0.2672 & 0.0021 & 0.0374 \\
200 & 0 & $m_{21}$ & 0.3318 & 0.3224 & -0.0094 & 0.0817 \\
200 & 0 & $m_{10}$ & 0.1540 & 0.1441 & -0.0099 & 0.0457 \\
200 & 0 & $m_{20}$ & 0.9318 & 1.0064 & 0.0746 & 0.3236 \\
200 & 1 & $m_{11}$ & 0.1929 & 0.1935 & 0.0006 & 0.0238 \\
200 & 1 & $m_{21}$ & 0.2762 & 0.2694 & -0.0068 & 0.0629 \\
200 & 1 & $m_{10}$ & 0.1095 & 0.1032 & -0.0063 & 0.0315 \\
200 & 1 & $m_{20}$ & 0.9429 & 1.0238 & 0.0810 & 0.3733 \\
500 & 0 & $m_{11}$ & 0.2652 & 0.2665 & 0.0013 & 0.0234 \\
500 & 0 & $m_{21}$ & 0.3318 & 0.3286 & -0.0032 & 0.0523 \\
500 & 0 & $m_{10}$ & 0.1540 & 0.1508 & -0.0032 & 0.0274 \\
500 & 0 & $m_{20}$ & 0.9318 & 0.9549 & 0.0231 & 0.1851 \\
500 & 1 & $m_{11}$ & 0.1929 & 0.1937 & 0.0009 & 0.0153 \\
500 & 1 & $m_{21}$ & 0.2762 & 0.2737 & -0.0025 & 0.0374 \\
500 & 1 & $m_{10}$ & 0.1095 & 0.1075 & -0.0021 & 0.0175 \\
500 & 1 & $m_{20}$ & 0.9429 & 0.9686 & 0.0258 & 0.1943 \\
\hline
\end{tabular}
\end{table}

\begin{table}[H]
\centering\small
\caption{Estimated marginal means under the full model}
\label{tab:additional_marginal}
\begin{tabular}{rrlrrrr}
\hline
$n$ & $x$ & Target & True & Mean & Bias & RMSE \\
\hline
\multicolumn{7}{l}{\textit{Canonical link}} \\
\hline
200 & 0 & $\mu_1$ & 0.2374 & 0.2371 & -0.0003 & 0.0239 \\
200 & 0 & $\mu_2$ & 0.4818 & 0.4822 & 0.0004 & 0.0431 \\
200 & 1 & $\mu_1$ & 0.1762 & 0.1759 & -0.0003 & 0.0170 \\
200 & 1 & $\mu_2$ & 0.4095 & 0.4081 & -0.0015 & 0.0341 \\
500 & 0 & $\mu_1$ & 0.2374 & 0.2377 & 0.0003 & 0.0152 \\
500 & 0 & $\mu_2$ & 0.4818 & 0.4817 & -0.0001 & 0.0269 \\
500 & 1 & $\mu_1$ & 0.1762 & 0.1765 & 0.0003 & 0.0110 \\
500 & 1 & $\mu_2$ & 0.4095 & 0.4092 & -0.0004 & 0.0212 \\
\hline
\multicolumn{7}{l}{\textit{Log link}} \\
\hline
200 & 0 & $\mu_1$ & 0.2374 & 0.2370 & -0.0003 & 0.0239 \\
200 & 0 & $\mu_2$ & 0.4818 & 0.4823 & 0.0005 & 0.0431 \\
200 & 1 & $\mu_1$ & 0.1762 & 0.1759 & -0.0003 & 0.0169 \\
200 & 1 & $\mu_2$ & 0.4095 & 0.4081 & -0.0014 & 0.0341 \\
500 & 0 & $\mu_1$ & 0.2374 & 0.2377 & 0.0003 & 0.0152 \\
500 & 0 & $\mu_2$ & 0.4818 & 0.4817 & -0.0001 & 0.0269 \\
500 & 1 & $\mu_1$ & 0.1762 & 0.1765 & 0.0003 & 0.0110 \\
500 & 1 & $\mu_2$ & 0.4095 & 0.4092 & -0.0004 & 0.0212 \\
\hline
\end{tabular}
\end{table}

\section{Conclusion}\label{sec-8}

In this paper, we developed two regression approaches for positive bivariate responses following the BBBE distribution. Its region-wise exponential-family representation provides a common basis for deriving moment and covariance identities and for constructing regression models through link functions, linear predictors, and likelihood-based estimation.

In the full-distribution approach, link functions connect the cumulant derivatives in a fixed reference region to linear predictors, while both regions contribute to the likelihood. These derivatives do not directly equal the marginal means, which are obtained as probability-weighted combinations of the regional derivatives. This relationship allows covariate effects on the marginal means to be examined. After maximum likelihood estimation of the regression coefficients, the rate parameter estimates are recovered through the inverse link functions and an explicit inverse mapping, with the admissible solution selected to ensure positive rates.

The conditional approach addresses situations in which interest lies in lifetimes conditional on a particular response ordering, namely $Y_1<Y_2$ or $Y_1>Y_2$. Normalizing the distribution within the corresponding region yields cumulant derivatives that coincide with the conditional means, allowing link functions to relate these means directly to linear predictors. The two approaches are therefore complementary, differing in the quantities linked to covariates and in the interpretation of the resulting coefficients.

The simulation study showed decreasing estimation variability with increasing sample size and small biases in the reconstructed means. Coverage was close to the nominal level for the conditional models, although some Wald intervals under the full log-link model were conservative. These findings support the implementation in the settings examined while indicating the need for further investigation of interval calibration.

The current framework concerns uncensored observations. Incorporating censoring would broaden its applicability in reliability and survival analysis. Further work may address diagnostics for the joint distribution, inference in small samples, and extensions to other bivariate distributions and multivariate responses.

\section*{Conflict of Interest}

On behalf of all authors, the corresponding author states that there is no conflict of interest.

  \bibliographystyle{elsarticle-harv} 
  \bibliography{reference}

@article{zhang2026inference,
  title={Inference for dependent competing risks model under m-cycle minimum ranked set sampling},
  author={Zhang, Jiaxin and Gui, Wenhao},
  journal={Journal of Applied Statistics},
  volume={53},
  number={8},
  pages={1463--1492},
  year={2026},
  publisher={Taylor \& Francis}
}

@article{samanta2024competing,
  title={On competing risk model under step-stress stage life testing},
  author={Samanta, Debashis and Kundu, Debasis},
  journal={Journal of Statistical Computation and Simulation},
  volume={94},
  number={4},
  pages={699--721},
  year={2024},
  publisher={Taylor \& Francis}
}

@article{ragab2019prognostics,
  title={Prognostics of multiple failure modes in rotating machinery using a pattern-based classifier and cumulative incidence functions},
  author={Ragab, Ahmed and Yacout, Soumaya and Ouali, Mohamed-Salah and Osman, Hany},
  journal={Journal of Intelligent Manufacturing},
  volume={30},
  number={1},
  pages={255--274},
  year={2019},
  publisher={Springer}
}

@article{samanta2021bayesian,
  title={Bayesian inference of a dependent competing risk data},
  author={Samanta, Debashis and Kundu, Debasis},
  journal={Journal of Statistical Computation and Simulation},
  volume={91},
  number={15},
  pages={3069--3086},
  year={2021},
  publisher={Taylor \& Francis}
}

@article{shen2018dependent,
  title={On the dependent competing risks using Marshall--Olkin bivariate Weibull model: Parameter estimation with different methods},
  author={Shen, Yan and Xu, Ancha},
  journal={Communications in Statistics-Theory and Methods},
  volume={47},
  number={22},
  pages={5558--5572},
  year={2018},
  publisher={Taylor \& Francis}
}

@article{wang2003bayesian,
  title={Bayesian analysis of bivariate competing risks models with covariates},
  author={Wang, Chen-Pin and Ghosh, Malay},
  journal={Journal of statistical planning and inference},
  volume={115},
  number={2},
  pages={441--459},
  year={2003},
  publisher={Elsevier}
}

@article{kumar2025robust,
  title={Robust and Efficient Estimation for the Block-Basu Bivariate Exponential Distribution: S. Kumar et al.},
  author={Kumar, Sanjay and Kundu, Debasis and Mitra, Sharmishtha},
  journal={Statistics and Computing},
  volume={35},
  number={5},
  pages={150},
  year={2025},
  publisher={Springer}
}

@article{javed2026logistic,
  title={A Logistic Regression Model to Evaluate Learning in a Semester System of Education Based on Classical and Objective Bayesian Approaches},
  author={Javed, Amara and Abbas, Nasir},
  journal={Journal of Statistical Theory and Applications},
  volume={25},
  number={1},
  pages={28},
  year={2026},
  publisher={Springer}
}

@article{dos2011bayesian,
  title={A Bayesian analysis for the Block and Basu bivariate exponential distribution in the presence of covariates and censored data},
  author={dos Santos, Carlos Aparecido and Achcar, Jorge Alberto},
  journal={Journal of Applied Statistics},
  volume={38},
  number={10},
  pages={2213--2223},
  year={2011},
  publisher={Taylor \& Francis}
}

@article{bonat2016multivariate,
  title={Multivariate covariance generalized linear models},
  author={Bonat, Wagner Hugo and J{\o}rgensen, Bent},
  journal={Journal of the Royal Statistical Society Series C: Applied Statistics},
  volume={65},
  number={5},
  pages={649--675},
  year={2016},
  publisher={Oxford University Press}
}

@article{yee1996vector,
  title={Vector generalized additive models},
  author={Yee, Thomas W and Wild, CJ},
  journal={Journal of the Royal Statistical Society Series B: Statistical Methodology},
  volume={58},
  number={3},
  pages={481--493},
  year={1996},
  publisher={Oxford University Press}
}

@article{smyth1989generalized,
  title={Generalized linear models with varying dispersion},
  author={Smyth, Gordon K},
  journal={Journal of the Royal Statistical Society: Series B (Methodological)},
  volume={51},
  number={1},
  pages={47--60},
  year={1989},
  publisher={Wiley Online Library}
}

@phdthesis{yousif2023comparative,
  title={A Comparative Analysis between Various Machine Learning Models and Generalized Linear Models},
  author={Yousif, Jan Mikael},
  year={2023},
  school={Stockholm University Stockholm, Sweden}
}

@article{lai2023extension,
  title={Extension of the glmm. hp package to zero-inflated generalized linear mixed models and multiple regression},
  author={Lai, Jiangshan and Zhu, Weijie and Cui, Dongfang and Mao, Lingfeng},
  journal={Journal of Plant Ecology},
  volume={16},
  number={6},
  pages={rtad038},
  year={2023},
  publisher={Oxford University Press UK}
}

@inproceedings{tsang2023post,
  title={Post-Selection Multiple Testing in Generalized Linear Models},
  author={Tsang, Ka Wai and Dai, Wei},
  booktitle={International Conference on Statistics, Applied Mathematics and Computing Science},
  pages={273--280},
  year={2023},
  organization={Springer}
}

@article{mathur2026parsimonious,
  title={Parsimonious Subset Selection for Generalized Linear Models with Biomedical Applications},
  author={Mathur, Anant and Liquet, Benoit and Muller, Samuel and Moka, Sarat},
  journal={arXiv preprint arXiv:2603.21952},
  year={2026}
}

@article{theus2026generalized,
  title={Generalized linear mode connectivity for transformers},
  author={Theus, Alexander and Cabodi, Alessandro and Anagnostidis, Sotiris and Orvieto, Antonio and Singh, Sidak Pal and Boeva, Valentina},
  journal={Advances in Neural Information Processing Systems},
  volume={38},
  pages={136197--136225},
  year={2026}
}

@article{knight2026fast,
  title={Fast and robust invariant generalized linear models},
  author={Knight, Parker and Jobe, Ndey Isatou and Duan, Rui},
  journal={Journal of Computational and Graphical Statistics},
  volume={35},
  number={3},
  pages={1064--1075},
  year={2026},
  publisher={Taylor \& Francis}
}

@article{correia2026verifying,
  title={Verifying the existence of maximum likelihood estimates for generalized linear models},
  author={Correia, Sergio and Guimar{\~a}es, Paulo and Zylkin, Tom},
  journal={Econometric Reviews},
  pages={1--53},
  year={2026},
  publisher={Taylor \& Francis}
}

@article{marshall1967multivariate,
  title={A multivariate exponential distribution},
  author={Marshall, Albert W and Olkin, Ingram},
  journal={Journal of the American Statistical Association},
  volume={62},
  number={317},
  pages={30--44},
  year={1967},
  publisher={Taylor \& Francis}
}

@article{block1974continuous,
  title={A continuous, bivariate exponential extension},
  author={Block, Henry W and Basu, AP},
  journal={Journal of the American Statistical Association},
  volume={69},
  number={348},
  pages={1031--1037},
  year={1974},
  publisher={Taylor \& Francis}
}

@article{liang1992multivariate,
  title={Multivariate regression analyses for categorical data},
  author={Liang, Kung-Yee and Zeger, Scott L and Qaqish, Bahjat},
  journal={Journal of the Royal Statistical Society: Series B (Methodological)},
  volume={54},
  number={1},
  pages={3--24},
  year={1992},
  publisher={Wiley Online Library}
}

@book{dobson2018introduction,
  title={An introduction to generalized linear models},
  author={Dobson, Annette J and Barnett, Adrian G},
  year={2018},
  publisher={Chapman and Hall/CRC}
}

@book{mccullagh1989generalized,
  author = {McCullagh, P. and Nelder, J.A.},
  publisher = {Chapman \& Hall},
  title = {Generalized Linear Models, Second Edition},
  year = 1989
}

@book{FahrmeirTutz1994,
  title     = {Multivariate Statistical Modelling Based on Generalized Linear Models},
  author    = {Fahrmeir, Ludwig and Tutz, Gerhard},
  publisher = {Springer},
  year      = {1994}
}

@inproceedings{neto2009bivariate,
  title={Bivariate generalized linear model for interval-valued variables},
  author={Neto, Eufrasio de A Lima and Cordeiro, Gauss M and de Carvalho, Francisco AT and dos Anjos, Ulisses U and da Costa, Abner G},
  booktitle={2009 International Joint Conference on Neural Networks},
  pages={2226--2229},
  year={2009},
  organization={IEEE}
}

@article{liang1986longitudinal,
  title={Longitudinal data analysis using generalized linear models},
  author={Liang, Kung-Yee and Zeger, Scott L},
  journal={Biometrika},
  volume={73},
  number={1},
  pages={13--22},
  year={1986},
  publisher={Oxford University Press}
}

@article{wedderburn1974quasi,
  title={Quasi-likelihood functions, generalized linear models, and the Gauss—Newton method},
  author={Wedderburn, Robert WM},
  journal={Biometrika},
  volume={61},
  pages={439--447},
  year={1974},
  publisher={Oxford University Press}
}

@article{mccullagh1983quasi,
  title={Quasi-likelihood functions},
  author={McCullagh, Peter and others},
  journal={The Annals of Statistics},
  volume={11},
  pages={59--67},
  year={1983},
  publisher={Institute of Mathematical Statistics}
}

@article{nelder1972generalized,
  title={Generalized linear models},
  author={Nelder, John Ashworth and Wedderburn, Robert WM},
  journal={Journal of the Royal Statistical Society: Series A (General)},
  volume={135},
  pages={370--384},
  year={1972},
  publisher={Wiley Online Library}
}

@article{iwasaki2005bivariate,
  title={A bivariate generalized linear model with an application to meteorological data analysis},
  author={Iwasaki, Masakazu and Tsubaki, Hiroe},
  journal={Statistical Methodology},
  volume={2},
  pages={175--190},
  year={2005},
  publisher={Elsevier}
}

@Manual{R,
  title        = {R: A language and environment for statistical
                  computing},
  author       = {{\textsf{R} Core Team}},
  organization = {R Foundation for Statistical Computing, Vienna, Austria},
  year         =2020,
}

 \section*{Appendix}
\appendix
\section{Proof of Lemma \ref{lema_pi}  }\label{proof:lema_pi}
Using \eqref{eq:BBdist} on the region $R_1:=\{0<y_1<y_2\}$,
\begin{align*}
P(Y_2>Y_1)
&=\int_0^\infty\int_{y_1}^\infty
\frac{\lambda_1(\lambda_2+\lambda_3)\lambda}{\lambda_1+\lambda_2}
\exp\!\big[-\lambda_1 y_1-(\lambda_2+\lambda_3)y_2\big]\;dy_2\,dy_1\\
&=\frac{\lambda_1(\lambda_2+\lambda_3)\lambda}{\lambda_1+\lambda_2}
\int_0^\infty e^{-\lambda_1 y_1}\left(\int_{y_1}^\infty e^{-(\lambda_2+\lambda_3)y_2}\,dy_2\right)dy_1\\
&=\frac{\lambda_1(\lambda_2+\lambda_3)\lambda}{\lambda_1+\lambda_2}
\int_0^\infty e^{-\lambda_1 y_1}\frac{e^{-(\lambda_2+\lambda_3)y_1}}{\lambda_2+\lambda_3}\,dy_1\\
&=\frac{\lambda_1\lambda}{\lambda_1+\lambda_2}\int_0^\infty e^{-\lambda y_1}\,dy_1
=\frac{\lambda_1}{\lambda_1+\lambda_2}.
\end{align*}
Hence for the region $R_0:=\{0<y_2<y_1\}$, $P(Y_1>Y_2)=1-P(Y_2>Y_1)=\dfrac{\lambda_2}{\lambda_1+\lambda_2}$. The proof of the expectation is straightforward by  using algebra.

\section{Proof of Lemma \ref{cond-moments} }\label{proof:cond-moments}
Using \eqref{eq:cond_y2gt},
\begin{align*}
\mathbb E_{R_1}(Y_1)
&=\int_0^\infty\int_{y_1}^\infty y_1\,P_{R_1}(y_1,y_2; \BS{\lambda}) dy_2\,dy_1\\
&=(\lambda_2+\lambda_3)\lambda\int_0^\infty y_1 e^{-\lambda_1 y_1}\,
\frac{e^{-(\lambda_2+\lambda_3)y_1}}{\lambda_2+\lambda_3}\,dy_1\\
&=\lambda\int_0^\infty y_1 e^{-\lambda y_1}\,dy_1
=\frac{1}{\lambda}.
\end{align*}
Similarly, using $\int_{y_1}^\infty y_2 e^{-a y_2}\,dy_2=(y_1+1/a)e^{-a y_1}/a$ with $a=\lambda_2+\lambda_3$,
\begin{align*}
\mathbb E_{R_1}(Y_2)
&=(\lambda_2+\lambda_3)\lambda\int_0^\infty e^{-\lambda_1 y_1}
\left(\int_{y_1}^\infty y_2 e^{-(\lambda_2+\lambda_3)y_2}\,dy_2\right)dy_1\\
&=(\lambda_2+\lambda_3)\lambda\int_0^\infty e^{-\lambda_1 y_1}
\left(\frac{y_1+1/(\lambda_2+\lambda_3)}{\lambda_2+\lambda_3}e^{-(\lambda_2+\lambda_3)y_1}\right)dy_1\\
&=\lambda\int_0^\infty y_1 e^{-\lambda y_1}\,dy_1
+\frac{\lambda}{\lambda_2+\lambda_3}\int_0^\infty e^{-\lambda y_1}\,dy_1\\
&=\frac{1}{\lambda}+\frac{1}{\lambda_2+\lambda_3}.
\end{align*}
For the second moments, we use the standard integral identities
\[
\int_0^\infty y^2 e^{-\lambda y}\,dy=\frac{2}{\lambda^3},
\qquad
\int_0^\infty y e^{-\lambda y}\,dy=\frac{1}{\lambda^2},
\]
from which the corresponding variances follow immediately. Moreover, on the event $\{Y_2>Y_1\}$, we may write $Y_2 = Y_1 + (Y_2-Y_1)$. Since, under the conditional distribution given $Y_2>Y_1$, the random variables $Y_1$ and $Y_2-Y_1$ are independent, it follows that
\[
\mathbb{E}_{R_1}\!\left[Y_1(Y_2-Y_1)\right]
=
\mathbb{E}_{R_1}(Y_1)\,
\mathbb{E}_{R_1}(Y_2-Y_1).
\]
Consequently,
\[
\operatorname{Cov}_{R_1}(Y_1,Y_2)
=
\operatorname{Cov}_{R_1}\bigl(Y_1,\;Y_1+(Y_2-Y_1)\bigr)
=
\operatorname{Var}_{R_1}(Y_1),
\]
which yields the desired covariance expression.
By symmetry, the corresponding results on $R_0$ and \eqref{eq:cond_y1gt} apply.

\section{Proof of Lemma \ref{marg-moments} }\label{proof:marg-moments}
From the exponential-family-like form of each density piece, we have
\[
\frac{\partial P_r(\BS{y};\boldsymbol{\theta})}{\partial\theta_j}
=
\left\{
y_j-
\frac{\partial b_r(\boldsymbol{\theta})}{\partial\theta_j}
\right\}
P_r(\BS{y};\boldsymbol{\theta}),
\qquad r=0,1.
\]
Substituting these derivatives into \eqref{eq:first-diff} and combining
the two regions using $\delta$ and $1-\delta$ give
\begin{align}
0
&=
\iint_{R_1\cup R_0}
\left\{
y_j
-
\delta\frac{\partial b_1(\boldsymbol{\theta})}{\partial\theta_j}
-
(1-\delta)
\frac{\partial b_0(\boldsymbol{\theta})}{\partial\theta_j}
\right\}
P_{\delta}(\BS{y};\boldsymbol{\theta})\,dy_1\,dy_2
\notag\\
&=
\mathbb{E}(Y_j)
-
\mathbb{E}\left\{
\delta\frac{\partial b_1(\boldsymbol{\theta})}{\partial\theta_j}
+
(1-\delta)
\frac{\partial b_0(\boldsymbol{\theta})}{\partial\theta_j}
\right\}.
\label{eq:first-diff-4}
\end{align}

Since the derivatives of $b_1(\boldsymbol{\theta})$ and
$b_0(\boldsymbol{\theta})$ are non-random for a fixed
$\boldsymbol{\theta}$, equation \eqref{eq:first-diff-4} gives
\begin{equation}
\mathbb{E}(Y_j)
=
\mathbb{E}(\delta)
\frac{\partial b_1(\boldsymbol{\theta})}{\partial\theta_j}
+
\mathbb{E}(1-\delta)
\frac{\partial b_0(\boldsymbol{\theta})}{\partial\theta_j},
\qquad j=1,2,3.
\label{eq:mean-component}
\end{equation}
Collecting the three component-wise identities in
\eqref{eq:mean-component} proves \eqref{eq:mean-vector-final}.
To derive the covariance matrix, write
\[
\mathbf T=\mathbf T(Y_1,Y_2),\qquad
b_\delta=\delta b_1+(1-\delta)b_0,
\qquad
\mathbf S=\mathbf T-\nabla b_\delta.
\]
Here, $\mathbf S$ is the score vector of the original density.
Differentiating the density twice gives
\[
\nabla^2 P(\BS y;\boldsymbol\theta)
=
\left\{
\mathbf S\mathbf S^\trp-\nabla^2b_\delta
\right\}P(\BS y;\boldsymbol\theta).
\]
Integrating over $R_1\cup R_0$ and differentiating the
normalization identity \eqref{eq:norm-split}
twice under the integral sign yields
\begin{equation}
\mathbb E(\mathbf S\mathbf S^\trp)
=
\mathbb E(\delta)\nabla^2b_1
+
\mathbb E(1-\delta)\nabla^2b_0.
\label{eq:score-second-moment}
\end{equation}
Since $\mathbb E(\mathbf S)=\mathbf 0$, this is the covariance
matrix of the score.

To obtain the covariance of $\mathbf T$, let
\[
p=p(\boldsymbol\theta)=\mathbb E(\delta),
\qquad
\mathbf d=\nabla b_1-\nabla b_0.
\]
Then $\nabla b_\delta=\nabla b_0+\delta\mathbf d$, and hence
\[
\mathbf T=\mathbf S+\nabla b_0+\delta\mathbf d.
\]
Moreover, differentiating the region probability gives
\[
\nabla p
=
\nabla\iint_{R_1}P_1(\BS y;\boldsymbol\theta)\,dy_1\,dy_2
=
\mathbb E(\delta\mathbf S)
=
\operatorname{Cov}(\mathbf S,\delta).
\]
Taking covariances in the preceding decomposition of $\mathbf T$,
using $\operatorname{Var}(\delta)=p(1-p)$ and
\eqref{eq:score-second-moment}, therefore gives
\begin{equation}
\begin{aligned}
\operatorname{Var}(\mathbf T)
={}&
p\nabla^2b_1+(1-p)\nabla^2b_0
+p(1-p)\mathbf d\mathbf d^\trp\\
&+
(\nabla p)\mathbf d^\trp
+\mathbf d(\nabla p)^\trp.
\end{aligned}
\label{eq:cov-matrix-corrected}
\end{equation}
The component-wise covariance and variance identities follow by
taking the corresponding entries of this matrix.

\end{document}